\documentclass[conference]{IEEEtran}
\makeatletter
\@ifundefined{labelindent}{}{\let\labelindent\relax}
\makeatother
\usepackage{enumitem}
\usepackage{layout}
\usepackage{amssymb}
\usepackage{pifont}
\newcommand{\cmark}{\ding{51}} 
\newcommand{\xmark}{\ding{55}} 
\ifCLASSINFOpdf
\else
\fi

\usepackage{xspace}
\usepackage{enumerate} 
\usepackage{amsmath}
\usepackage{pgfplots}
\usepackage{multirow}
\usepackage{algorithm}
\usepackage{algpseudocode}
\usepackage{booktabs} 
\usepackage{graphicx} 
\usepackage[caption=false,font=footnotesize]{subfig} 
\usepackage{gensymb}
\usepackage{svg}
\usepackage{graphicx}
 \usepackage{subfig} 
 \usepackage{tikz}
 \usepackage{array}
\usetikzlibrary{calc, intersections, angles, quotes, decorations.markings}
\usetikzlibrary{shapes.geometric, arrows.meta, positioning, fit, calc}
\usetikzlibrary{decorations.markings}
\usepackage{subfig}
 \usetikzlibrary{shapes.geometric, arrows.meta, positioning, fit, calc}
 \usepackage[hidelinks]{hyperref}

\begin{document}

%
\captionsetup{font=small}
\title{Seeing is Deceiving: Mirror-Based LiDAR Spoofing for Autonomous Vehicle Deception}


%
\author{\IEEEauthorblockN{Selma Yahia\IEEEauthorrefmark{1},
Ildi Alla\IEEEauthorrefmark{1},
Girija Bangalore Mohan\IEEEauthorrefmark{2}, 
Daniel Rau\IEEEauthorrefmark{2},
Mridula Singh\IEEEauthorrefmark{3} and 
Valeria Loscri\IEEEauthorrefmark{1}}

\IEEEauthorblockA{\IEEEauthorrefmark{1}Inria Lille-Nord Europe, France}
\IEEEauthorblockA{\IEEEauthorrefmark{2}University of Applied Science Saarland - htw saar, Saarbrücken, Germany}
\IEEEauthorblockA{\IEEEauthorrefmark{3}CISPA Helmholtz Center for Information Security, Saarbrücken, Germany}}



\maketitle

\begin{abstract}

Autonomous vehicles (AVs) rely heavily on LiDAR sensors for accurate 3D perception of their surroundings. However, the physical principles underlying LiDAR make it inherently susceptible to optical manipulation. Prior work on LiDAR spoofing has focused primarily on active attacks, such as injecting fake laser pulses or tampering with sensor electronics. Some works have also investigated passive attacks involving expensive, custom-fabricated artifacts like 3D-printed objects or tailored road patterns.
In this paper, we introduce a novel class of low-cost passive LiDAR spoofing attacks that exploit mirror-like surfaces to inject or remove objects from the AV’s perception. By using planar mirrors to redirect LiDAR beams, our approach introduces a practical and stealthy threat -- one that require no electronics or custom fabrication and can be easily deploy in real-world settings. 
We introduce a comprehensive threat framework that encompasses two adversarial goals: Object Addition Attacks (OAA), which create phantom obstacles, and Object Removal Attacks (ORA), which conceal real objects from the AV's perception. We formalize the physical underpinnings of mirror-induced deception using geometric optics and validate our models through controlled real-world experiments. We demonstrated the impact of these attacks on autonomous driving stack (Autoware) by disrupting object detection, contaminating occupancy grids, and triggering hazardous driving behaviors.
We designed a simulation-based attack framework within the CARLA simulator, allowing safe and repeatable testing of various attack configurations using our empirical models. 
Our findings reveal a critical blind spot in existing AV systems that severely disrupt LiDAR perception without requiring any active transmission or hardware compromise. 
In addition, we provide discussion of potential defense approaches, while pointing out their limitations under real-world constraints.

\end{abstract}


%

\section{Introduction and Motivation}

The rapid evolution and deployment of Autonomous Driving (AD) technologies on public roads is fundamentally reliant on sophisticated sensor arrays that enable vehicles to accurately perceive their surroundings. Among these, Light Detection and Ranging (LiDAR) stands as a primary sensor, providing high-resolution 3D environmental information crucial for safe navigation and informed decision-making. The integrity and reliability of the data generated by LiDAR systems are therefore paramount, forming the bedrock upon which autonomous vehicles build their understanding of the world and execute critical maneuvers \cite{zhang2024lidar,leong2024lidar}.
Despite their advanced capabilities, LiDAR systems are inherently susceptible to deception stemming from various optical phenomena~\cite{bhupathiraju2023emi, li2024detection,suzuki2024wip}. An adversary can introduce erroneous data into the LiDAR's perception, manifesting as either phantom obstacles or, more dangerously, the concealment of real hazards \cite{Yang2011}. Such inconsistencies can lead to errors in localization, mapping, and navigation for mobile robots and 3D reconstruction \cite{Zhao2020}.

LiDAR spoofing attacks typically rely on active signal injection using laser emitters, which require precise synchronization and a direct line of sight to the victim sensor \cite{petit2015remote,sato2023lidar,jin2023pla,shin2017illusion,wang2023adversarial}, or on signal jamming, which requires high optical power to overwhelm the sensor with noise \cite{liu2021seeing,yan2016can}.
Passive approaches have been explored as well, but they typically depend on specially crafted physical artifacts such as 3D-printed objects or printed road patterns tailored to exploit perception models \cite{tu2020physically,cao2021invisible,yang2021robust}. While these methods target model-specific vulnerabilities using artificial inputs, they largely overlook a more fundamental avenue of deception: manipulation of LiDAR signal path itself.


Classic works in robotics \cite{Yang2011,damodaran2023experimental} have long recognized the challenges that reflective surfaces, such as mirrors, pose for LiDAR-based mapping. There also exists work that focuses on enhancing robustness in the presence of such surfaces \cite{vega2024slam2ref,koch2017detection,henley2023detection}.
However, the intentional use of these reflective surfaces remain largely unexplored in the context of adversarial perception research.  
Table~\ref{tab:comparison} provide a comparative overview of prior LiDAR deception capabilities, highlighting the absence of passive, mirror-based approaches in adversarial research. This reveals a critical gap in our understanding of a low-cost, passive threats that could be used intentionally to deceive LiDAR perception in real-world environments.

Addressing this gap, we show how the insertion of planar mirrors in the environment is a vector of powerful attacks, generating misleading perception of the real environmental conditions. Based on our analysis, we identify the following four key research limitations that currently hinder the demonstration and evaluation of mirror-based LiDAR spoofing attacks in practical autonomous driving scenarios:


\begin{table*}[t]
\centering
\caption{Comparison of Mirror-Based LiDAR Attack Capabilities with Prior Work}
\label{tab:comparison}
\begin{tabular}{|p{0.5cm}|p{0.7cm}|p{0.7cm}|p{1cm}|p{1cm}|p{1.5cm}|p{1.8cm}|p{3.2cm}|p{2.1cm}|p{1.5cm}|}
\hline
\textbf{Work} & \textbf{Mirror\-Based} & \textbf{Passive} & \textbf{Object Addition} & \textbf{Object Removal} & \textbf{Geometric Modeling} & \textbf{Simulation} & \textbf{Real-World Validation} & \textbf{Real-World AV Stack Eval.} & \textbf{Defense Proposal} \\
\hline
\textbf{Ours} & \cmark & \cmark & \cmark & \cmark & \cmark & \cmark (CARLA) & \cmark  (Outdoor, on-vehicle test) & \cmark (Autoware on exp. vehicle)  & \cmark \\
\hline
 \cite{tu2020physically} & \xmark & \cmark & \cmark & \xmark & \xmark & \xmark & \cmark (Indoor, basic test with VLP) & \xmark & \xmark\\
\hline
\cite{cao2021invisible} & \xmark & \cmark & \cmark & \xmark & \xmark & \xmark & \cmark (Outdoor) & \xmark & \xmark \\
\hline
\cite{yang2021robust} & \xmark & \cmark & \cmark & \xmark & \cmark & \cmark (LGSVL) & \cmark(Outdoor, perception-level) & \xmark & \cmark \\
\hline
\cite{kobayashi2024wip} & \cmark & \cmark & \xmark & \cmark & \xmark & \cmark (AWSIM) &  \xmark & \xmark & \cmark  \\
\hline
 \cite{kobayashiinvisible} & \cmark & \cmark & \xmark & \cmark & \xmark & \cmark (AWSIM) & \cmark(Outdoor, basic test with VLP) & \xmark & \cmark \\
\hline
 \cite{zhu2024ae} & \xmark & \cmark & \xmark & \xmark & \cmark  &  \cmark (SVL) & \cmark (Outdoor)& \xmark & \cmark \\
\hline
\cite{cao2019adversarial} & \xmark & \xmark & \cmark & \xmark & \xmark  & \cmark (Baidu Apollo)  & \cmark (Basic indoor test) & \xmark & \cmark \\
\hline
 \cite{sun2020towards} & \xmark & \xmark & \cmark & \xmark & \cmark & \cmark (On dataset) & \cmark(Indoor, basic test with VLP) & \xmark & \cmark \\
\hline
 \cite{zhu2021can} & \xmark & \cmark & \xmark & \cmark & \cmark & \cmark (On dataset) &\cmark(Outdoor, perception-level with drones) & \xmark & \cmark \\
\hline
\end{tabular}
\end{table*}

\textbf{Lack of Formalization of Mirror-Induced Attack Vectors:}~Recent work such as the ``Shadow Hack" attack \cite{kobayashi2024wip,kobayashiinvisible} has shown how reflective materials like aluminum mats can be used to remove LiDAR points and deceive object detection systems in autonomous vehicles. However, these studies have focused narrowly on point removal and have not conceptualized mirror-based deception as a distinct or formal class of adversarial attacks. More critically, they overlook the broader spectrum of threats posed by reflective surfaces, particularly the deliberate injection of phantom objects via redirected LiDAR beams.  Additional research \cite{zhu2024ae,cao2019adversarial,zhu2021can} examined adversarial materials such as shiny road signs or reflective cardboard cutouts designed to trigger classification or segmentation errors. These approaches, however,  did not exploit the unique optical properties of mirrors, specifically, their ability to precisely redirect incident LiDAR pulses and create spatially consistent, ghost-like artifacts in empty regions of space.

\textbf{Lack of Scientific Modeling for Mirror Effects:}~Mirror-based deception follows the laws of geometric optics, yet previous research has not established any mathematical or empirical models to describe how variables such as angle, distance, and surface area influence point cloud distortions. In the absence of such models, it becomes difficult to predict or simulate mirror-induced artifacts accurately, which hinders progress in both attack research and the development of effective defensive strategies.

\textbf{Lack of Simulation Frameworks for Safe, Repeatable Testing:}~Physical attack demonstrations are constrained by cost, safety, and experimental scalability. Although some recent frameworks address general LiDAR modeling \cite{manivasagam2020lidarsim,manivasagam2023towards}, none focus on simulating realistic mirror-induced artifacts in autonomous vehicle environments. The absence of such targeted simulation capabilities hinders repeatable experimentation, robust evaluation of perception systems, and large-scale training of defense mechanisms.

\textbf{Lack of End-to-End Evaluation on Full AV Stacks:}~Most prior studies evaluated attacks at the level of perception models or raw point clouds, without integrating the effects into a complete AV software stack \cite{cao2019adversarial,sun2020towards,zhu2024ae}. As a result, the broader implications of mirror-based attacks—such as false emergency braking, steering faults, or unsafe lane decisions—remain unexamined. There is a need to demonstrate how such attacks propagate through planning and control layers of real-world AV software.

Driven by these blind spots, this work investigates the following central question:
\textit{How can an adversary exploit ordinary reflective surfaces to construct LiDAR perception illusions, and what are the safety implications of these illusions for real-world autonomous vehicle systems?}

To answer this, we conducted the first in-depth examination of passive, mirror-based LiDAR deception in an automotive context. Our main contribution is an end-to-end analysis methodology that combines real-world experimentation, empirical modeling, and simulation-based validation to assess system-level vulnerabilities. We evaluate the full-stack impact of mirror-induced artifacts through real experiments using an autonomous vehicle operated by Autoware, showing how these deceptive point clouds are interpreted as legitimate environmental features and propagate through the perception, planning, and control pipelines. This results in unsafe driving behaviors, including false emergency braking. To simulate this attack behavior, we designed and implemented a CARLA-based simulation framework that enables the real-time injection of mirror-induced LiDAR artifacts. Our framework allows for safe, repeatable, and scalable testing of mirror attack scenarios under various traffic and environmental conditions, offering a controlled environment to study their effects and support future defense development.

In summary, we make the following key contributions:
\begin{itemize}[noitemsep]
 \item We formulate two mirror-based LiDAR attack scenarios: Object Addition Attack (OAA), which introduces \textit{phantom obstacles} by reflecting LiDAR pulses onto benign surfaces, and Object Removal Attack (ORA), which causes \textit{stealth occlusions} by redirecting pulses away from real objects. 
 \item We derive geometric optics models that characterize the formation of mirror-induced LiDAR artifacts via multi-path reflections. These models quantify how mirror angle, distance, and size influence the presence or absence of points in the LiDAR frame. We validate the models through controlled real-world experiments using a commercial Ouster LiDAR, showing how physical parameters systematically shape the resulting point clouds.
 
\item  We implement a real-time, simulation-based attack injection framework in CARLA to support safe, repeatable, and scalable testing of diverse mirror attack scenarios. The framework leverages our geometric models to synthesize realistic LiDAR artifacts during simulation.

 \item We assess the \textit{end-to-end impact} of mirror-based attacks through real-world experiments using an autonomous vehicle running the full Autoware driving stack. Our evaluation traces how mirror-induced LiDAR artifacts propagate through perception, planning, and control modules. We show that these artifacts can corrupt occupancy maps, cause false object detections, and trigger unsafe decisions such as emergency braking.
 

 \item We explore initial defense strategies, including thermal imaging, light fingerprinting, and multi-sensor fusion, to detect and mitigate mirror-based attacks.
\end{itemize}

\noindent\textbf{Project Website.}~Demonstration videos and results from both physical experiments and CARLA simulations are available at: \url{https://lidar-attack.github.io/lidar-research/}. The site highlights how mirror parameters influence fake point generation and AV perception.

\section{Background: Understanding LiDAR Technology}
\label{Section: Background}

\subsection{LiDAR Operation and Reflection Physics}

LiDAR sensors are foundational components of modern AV perception systems, providing dense 3D representations of the environment across a wide range of lighting and weather conditions. A typical LiDAR system emits pulses of near-infrared laser light and measures the round-trip time between emission and return to compute the distance to a surface using the time-of-flight principle. Estimated distance  $d = \frac{c \times \Delta t}{2}$, where $c$ is the speed of light, and $\Delta t$ is the measured time interval between emission and detection.



Contemporary LiDAR systems scan the environment using rotating or solid-state emitter–receiver arrays with fixed angular resolution in azimuth and elevation. The resulting data forms a 3D \textit{point cloud}, $P = { (x_i, y_i, z_i, r_i) }_{i=1}^N$, where $(x_i, y_i, z_i)$ is the 3D location and $r_i$ the return intensity. This return value is influenced by factors such as surface reflectivity, distance, and angle of incidence. Consequently, point cloud fidelity is tightly linked to how beams interact with physical surfaces.

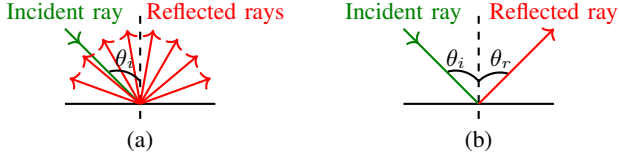
\begin{figure}[t]
    \centering
    \begin{tikzpicture}[scale=1.0]
        \tikzset{every node/.style={font=\small}}
        \tikzset{every path/.style={thick}}

        \definecolor{lightgreen}{rgb}{0, 0.5, 0}
        \definecolor{lightred}{rgb}{1, 0, 0}       

        \coordinate (b) at (0, 0);       
        \coordinate (a) at (4.5, 0);     

        \begin{scope}[shift={(b)}]
            \node at (0, -0.5){(a)}; 
            \draw[thick] (-1, 0) -- (1, 0); 
            \draw[dashed] (0, -0.2) -- (0, 1.2); 
            \draw[lightgreen, postaction={decorate}, decoration={markings, mark=at position 0.2 with {\arrow{>}}}] (-1, 1) -- (0, 0); 
            \foreach \angle in {20, 40, 60, 80, 100, 120, 140, 160} {
                \draw[->, thick, lightred] (0, 0) -- ({cos(\angle)}, {sin(\angle)});
            }
            \node[lightgreen] at (-1, 1.2) {Incident ray};
            \node[lightred] at (1, 1.2) {Reflected rays};
            \node at (-0.2, 0.6) {\(\theta_i\)};
            \draw[thick] (0, 0.3) arc[start angle=40,end angle=105,radius=0.4]; 
        \end{scope}

        \begin{scope}[shift={(a)}]
            \node at (0, -0.5) {(b)};
            \draw[thick] (-1, 0) -- (1, 0); 
            \draw[dashed] (0, -0.2) -- (0, 1.2); 
            \draw[lightgreen, postaction={decorate}, decoration={markings, mark=at position 0.2 with {\arrow{>}}}] (-1, 1) -- (0, 0); 
            \draw[->, thick, lightred] (0, 0) -- (1, 1); 
            \node[lightgreen] at (-1, 1.2) {Incident ray};
            \node[lightred] at (1, 1.2) {Reflected ray};
            \node at (-0.3, 0.6) {\(\theta_i\)};
            \node at (0.3, 0.6) {\(\theta_r\)};
            \draw[thick] (0, 0.3) arc[start angle=40,end angle=105,radius=0.4]; 
            \draw[thick] (0, 0.3) arc[start angle=135,end angle=75,radius=0.4]; 
        \end{scope}
    \end{tikzpicture}
    \captionsetup{font=small}
    \caption{Types of reflection in LiDAR sensing. (a) Diffuse reflection: incident rays scatter in many directions, yielding returns from rough surfaces across broad angles. (b) Specular reflection: a smooth mirror produces a coherent reflected ray at angle $\theta_r$ equal to incidence $\theta_i$.
    }
    \label{fig:reflection_types}
\end{figure}

Two main types of surface interaction determine LiDAR returns: \textit{diffuse} and \textit{specular} reflection, as shown in Figure~\ref{fig:reflection_types}. Diffuse reflection occurs when the surface is rough, scattering the return energy broadly and allowing detection from multiple angles. Specular reflection occurs on smooth, mirror-like surfaces such as glass windows, wet road surfaces, and glossy vehicle bodies, and follows the classic law of reflection:
\begin{equation}\small
    \vec{v}_{\text{ref}} = \vec{v}_{\text{in}} - 2 \left( \vec{v}_{\text{in}} \cdot \vec{n} \right) \vec{n},
    \label{eq:specular_reflection}
\end{equation}
where $\vec{v}_{\text{in}}$ is the incoming ray and $\vec{n}$ is the surface normal. This behavior is the primary source of complex measurement artifacts in LiDAR data.

\subsection{LiDAR-Mirror Interactions: Measurement Artifacts}
\label{Section: BackgroundB}

\begin{figure}[t]
    \centering

    \subfloat[]{%
      \begin{tikzpicture}[scale=0.6]
        \tikzset{lidar/.style={circle, draw, fill=gray!30, minimum size=6mm}}
        \tikzset{object/.style={rectangle, draw, fill=blue!20, minimum height=2cm, minimum width=0.2cm}}
        \tikzset{mirror/.style={draw=blue, ultra thick}}
        \tikzset{ray/.style={-stealth, red, thick, postaction={decorate}, decoration={markings, mark=at position 0.6 with {\arrow{>}}}}}
        \node[lidar] (L) at (0,0) {L};
        \node at (L.south) [below=1mm] {LiDAR};
        \node[object] (O) at (3.5,0) {};
        \node at (O.south) [below=1mm] {Object};
        \draw[mirror] (2,-1.5) -- (2,1);
        \node at (2, 1.8) {Mirror};
        \draw[ray] (L) -- (2, 0.5);
        \draw[ray, dashed] (2, 0.5) -- (3.5, 2);
        \node at (3.5, 2.2) [font=\tiny] {No return};
        \draw[gray, dashed, -stealth] (L) -- (O);
      \end{tikzpicture}
      \label{fig:hiding_sub}
    }\hfill
    \subfloat[]{%
      \begin{tikzpicture}[scale=0.65]
        \tikzset{lidar/.style={circle, draw, fill=gray!30, minimum size=6mm}}
        \tikzset{object/.style={circle, draw, fill=green!30, minimum size=6mm}}
        \tikzset{mirror/.style={draw=blue, ultra thick, name path=mirror}}
        \tikzset{actual_ray/.style={-stealth, red, thick, postaction={decorate}, decoration={markings, mark=at position 0.4 with {\arrow{<}}, mark=at position 0.7 with {\arrow{>}}}}}
        \tikzset{perceived_ray/.style={-stealth, red, thick, dashed}}
        \coordinate (L) at (-3, 1.5);
        \coordinate (S) at (-2.5, -1);
        \coordinate (M_top) at (-1, 2); \coordinate (M_bottom) at (-1, -2);
        \path[mirror] (M_top) -- (M_bottom);
        \coordinate (F) at (0.5, -1);
        \path[name path=perceived_path] (L) -- (F);
        \path[name intersections={of=mirror and perceived_path, by=M}];
        \draw[mirror] (M_top) -- (M_bottom);
        \node[lidar, label=left:\tiny{$\mathbf{P}_L$}] at (L) {};
        \node[object, label=below:\tiny{$\mathbf{P}_S$}] at (S) {};
        \fill[red!80!black] (F) circle (1.5pt) node[right, font=\tiny]{$\mathbf{P}_F$};
        \draw[perceived_ray] (L) -- (F) node[midway, below, sloped, font=\tiny] {$d_{LM} + d_{MS}$};
        \draw[actual_ray] (L) -- (M) node[midway, above, sloped, font=\tiny] {$d_{LM}$};
        \draw[actual_ray] (M) -- (S) node[midway, below, sloped, font=\tiny] {$d_{MS}$};
      \end{tikzpicture}
      \label{fig:virtual_sub}
    }

    \caption{Primary LiDAR measurement artifacts from specular reflections. (a) Data omission occurs when a beam is reflected away from the sensor. (b) Data fabrication occurs when a beam reflects off a secondary object, creating a virtual point behind the mirror.}
    \label{fig:mirror_artifacts_combined}
\end{figure}
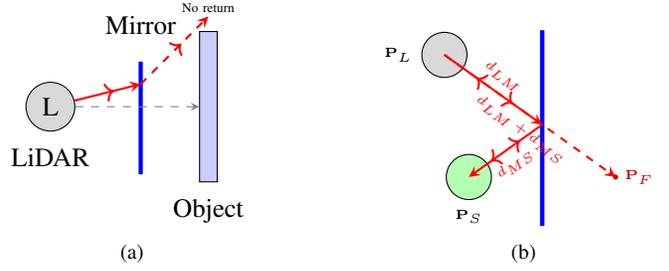

When a LiDAR beam strikes a specular surface, its path is redirected according to the law of reflection. This can corrupt the resulting point cloud by producing two main types of artifacts, as shown in Figure~\ref{fig:mirror_artifacts_combined}: \textbf{data omission}, where real objects are hidden, and \textbf{data fabrication}, where non-existent “virtual” objects appear.

An example of data omission is shown in Figure~\ref{fig:mirror_artifacts_combined}~(a). This occurs when a mirror deflects a LiDAR beam away from the sensor, preventing a return signal. A beam that would normally strike a real object ($\mathbf{P}_{O}$) instead reflects along a path that never returns, causing the LiDAR to record empty space. The object is effectively removed from the point cloud, creating a blind spot that can compromise navigation and obstacle avoidance.

Conversely, a more deceptive artifact involves the creation of virtual points, illustrated in Figure~\ref{fig:mirror_artifacts_combined}~(b). This occurs when a LiDAR beam is redirected by a mirror, strikes a secondary object ($\mathbf{P}_{S}$), and then reflects back to the sensor via the same mirror. The light follows a multi-hop path, but the LiDAR system assumes a direct line of sight and misinterprets the total travel time, resulting in an incorrect depth estimate:
\begin{equation}
    d_{\text{measured}} = \frac{d_{\text{total}}}{2} = d_{LM} + d_{MS},
\end{equation}
where $d_{LM}$ is the distance from the LiDAR to the mirror, and $d_{MS}$ is the distance from the mirror to the secondary object. The system projects the total travel distance along the original emission direction, assuming a direct path. As a result, the reflected object appears at a false location, $\mathbf{P}_F$, the geometric mirror image of the actual object. This creates phantom obstacles in the point cloud that may be mistaken for real, compromising object detection and scene understanding.

\section{Threat Model}
\label{sec:threat_model}
We consider an adversary aiming to compromise the integrity of a victim AV's perception system by manipulating its LiDAR sensor data. The attack is executed entirely in the physical domain by simply placing mirrors in the environment. This approach constitutes a form of physical-world adversarial example, where the environment itself is altered to deceive the sensor. Unlike attacks that require electronic intrusion or signal injection, our threat model focuses on a non-invasive adversary who subverts the LiDAR's operation by redirecting the vehicle's own emitted laser pulses.
\subsection{Adversary Goals}
The adversary's primary objective is to introduce physically plausible but semantically deceptive artifacts into the AV's LiDAR point cloud, leading to unsafe or disruptive vehicle behavior. We define two principal attacks:

\noindent $\bullet$  \textbf{Object Addition Attack (OAA):}~The adversary's goal is to create a \textit{false positive} by fabricating a ``ghost" obstacle in the AV's path. By making a non-existent object appear in the LiDAR data, the attacker aims to trigger an inappropriate defensive response, such as sudden emergency braking in flowing traffic or an unnecessary evasive maneuver into an adjacent lane.

\noindent $\bullet$  \textbf{Object Removal Attack (ORA):}~The adversary's goal is to create a \textit{false negative} by making a real, physical obstacle invisible to the LiDAR sensor. By deflecting LiDAR beams away from a genuine hazard, the attacker aims to cause the AV to fail to detect a threat, potentially leading to a collision.
 
Crucially, both attacks are designed to be covert. They do not introduce obvious noise or degradation but instead produce structured, plausible point clouds that a perception system would likely interpret as a normal, albeit dangerous, scene.


\subsection{Adversary Model}
We consider a \textit{passive, external adversary} operating entirely in the physical domain. The capabilities and constraints of the adversary are as follows:

\noindent $\bullet$ \textbf{Black-Box System Access:}~The adversary has no internal access to the AV’s software, sensor data, or control systems, and does not rely on hacking or software exploits. The attack is non-invasive and operates without feedback from the vehicle’s perception module, relying only on observable physical reactions such as braking or swerving.

\noindent $\bullet$ \textbf{Passive, Reflection-Based Attack Vector:}~The adversary does not employ any electronic emitters. Unlike spoofing or jamming attacks that actively generate counterfeit signals \cite{sato2025realism, shin2017illusion}, this attack is carried out by passively redirecting the victim vehicle's own LiDAR pulses.  This ensures the received signals are physically legitimate, yet geometrically deceptive.

\noindent $\bullet$ \textbf{General Environmental and System Knowledge:}~The adversary is assumed to have general knowledge of the target environment, such as the layout of the road or intersection segment the AV is expected to travel. They also possess high-level knowledge of typical LiDAR sensor technology and common sensor mounting locations (e.g., rooftop, bumper), but do not require precise calibration parameters of the specific target vehicle.

\noindent $\bullet$ \textbf{Physical Manipulation of the Environment:}~The adversary can place one or more planar mirrors in the environment before the AV arrives. These could be portable, moderate-sized mirrors (e.g., tens of centimeters per side) that are freestanding or camouflaged as part of the existing infrastructure, such as being affixed to a utility box, hidden in a construction sign, or integrated into roadside art.

\begin{figure}[t]
\centering
    \includegraphics[width=\linewidth]{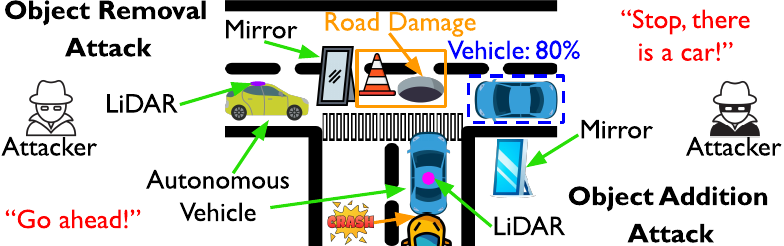}
\captionsetup{font=small}    
\caption{Mirror-Based LiDAR Spoofing Scenarios:  In the ORA, a mirror redirects LiDAR beams away from a traffic cone used to signal road damage, causing the AV to perceive a clear path where a hazard exists. In the OAA, a mirror reflects LiDAR pulses to create a phantom obstacle, leading to a false positive detection.}
\label{fig:threat model}
\end{figure}

\subsection{Representative Attack Scenarios}\label{attack_scenarios}

We describe two realistic deployment scenarios corresponding to the ORA and OAA attacks, as illustrated in Figure~\ref{fig:threat model}, to concretely ground the threat model. Each scenario is chosen to be plausible in an urban setting without requiring improbable precision or insider access.

\noindent $\bullet$ \textbf{Scenario 1:~Turn-Blocking via Object Addition (OAA).}~An adversary targets an autonomous vehicle approaching an urban intersection to execute a right or left turn (see Figure~\ref{fig:threat model}). A planar mirror is positioned on the corner sidewalk, angled to intercept LiDAR beams emitted as the vehicle nears the turn.  Some pulses strike the mirror and are redirected onto the vehicle’s own side body, creating a multi-hop return path that forms a phantom echo (see Section~\ref{Section: BackgroundB}). 
This false return appears geometrically aligned with the turning path and is misinterpreted by the vehicle’s perception system as a real obstacle, potentially causing it to brake or swerve unnecessarily.



\noindent $\bullet$ \textbf{Scenario 2:~``Paving Over" an Obstacle via Object Removal (ORA).}~An attacker targets an autonomous vehicle traveling along a road segment that contains a real obstacle, such as a construction barrier or roadwork equipment (see Figure~\ref{fig:threat model}). A planar mirror is placed at the roadside, crossing the vehicle’s line-of-sight and angled to reflect the sky or ground. This setup diverts outgoing LiDAR pulses away from the actual obstacle. The redirected beams return measurements corresponding to an unobstructed region,  leading the system to infer that the space ahead is clear and drivable. 

\section{Outdoor Experimental Campaign}
\label{Outdoor_exp}

This section describes the preliminary real-world experiments we conducted to analyze the physical behavior of mirror-induced LiDAR artifacts under ORA and OAA attacks. All experiments were performed using a consistent sensor platform on a test vehicle, and were designed to capture raw LiDAR data under realistic outdoor configurations. We focus on how mirror placement impacts the LiDAR’s point cloud geometry and  detection results.


\subsection{Platform and Sensor Configuration}

All experiments utilized a test vehicle outfitted with an \textit{Ouster OS1-128} LiDAR sensor, roof-mounted at a height of approximately 2.2\,m. The LiDAR operated at 10\,Hz, producing full 360° scans with 128 vertical channels (a typical high-resolution sensor for AV research) \cite{ouster2024os1}. Data was recorded using ROS\,2, logging raw LiDAR packets for precise offline analysis. We used Autoware’s runtime visualization to monitor detection outputs for sanity checks during trials.

The mirrors used were optically flat, high-reflectivity glass panels with silver backing. Each mirror was mounted on adjustable stands with angular protractors to precisely control tilt angles. We had mirrors of various sizes: the primary ones measured about 60\,cm$\times$40\,cm (for ORA tests) and we also assembled a modular array of 30\,cm$\times$30\,cm tiles for larger reflective surfaces in OAA tests, with a total cost of around 60~USD. The mirror surfaces were kept clean to maintain high specular reflectance.

Testing was done in a closed parking lot and adjacent roadway on our campus, under mostly clear weather (daylight, dry pavement). We avoided very windy conditions for stability of mirror stands. The vehicle was driven manually along predefined paths approaching the mirror setups, at low speeds (under 8\,km/h) to replicate slow approach scenarios (e.g., coming to a stop or starting from a stop). This allowed us to capture the full effect of the mirror over multiple LiDAR frames as the geometry between car and mirror changed.

\subsection{Object Removal Attack Trials (ORA)}

\begin{figure}[t]
\centering
    \includegraphics[width=\linewidth]{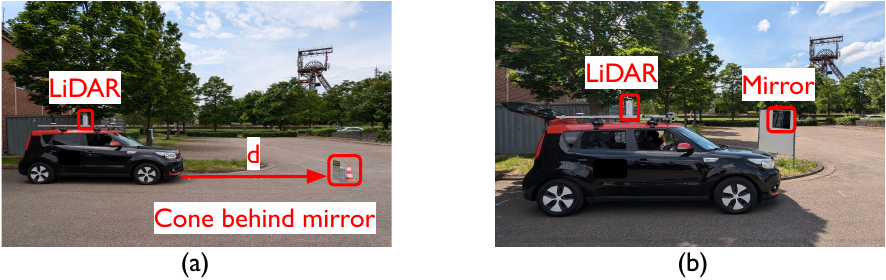}
\captionsetup{font=small}    
\caption{Experimental setups for the two mirror-based LiDAR attacks. \textbf{(a)} The ORA, where a planar mirror is positioned to obscure a traffic cone from the LiDAR's line-of-sight. \textbf{(b)} The OAA, where a mirror is angled to redirect LiDAR pulses toward a roadside structure, creating a phantom obstacle in the vehicle's path.}
\label{fig:Real_ORA_OAA}
\end{figure}

The first set of experiments targeted the ORA scenario, where the attacker attempts to \textit{conceal a real object} from LiDAR perception by redirecting outgoing pulses. In this setup, we used a traffic cone (50\ cm tall, base diameter 29\,cm) as the test obstacle (see Figure \ref{fig:Real_ORA_OAA}~(a)). The cone was placed at a fixed distance of 4\,m directly in front of the vehicle, on level ground. We then positioned a planar mirror (60\,cm$\times$40\,cm) between the LiDAR and the cone, such that it could \textit{obscure the line-of-sight} to the cone. The mirror’s bottom edge was just above the ground, fully covering the vertical span of the cone as seen by the LiDAR. We varied the mirror’s tilt angle relative to the LiDAR’s forward axis: tested angles were 0\degree \,(mirror facing straight toward the LiDAR), 15\degree, 30$^\circ$, and $45^\circ$. Each configuration was executed twice, with each run lasting 10 seconds and capturing approximately 100 LiDAR frames, which allowed us to average out noise and account for any slight movements. The vehicle remained stationary in these trials to isolate the effect of the mirror from any motion. The LiDAR and mirror positions were fixed for each angle.

\begin{figure}[t]
\centering
    \includegraphics[width=\linewidth]{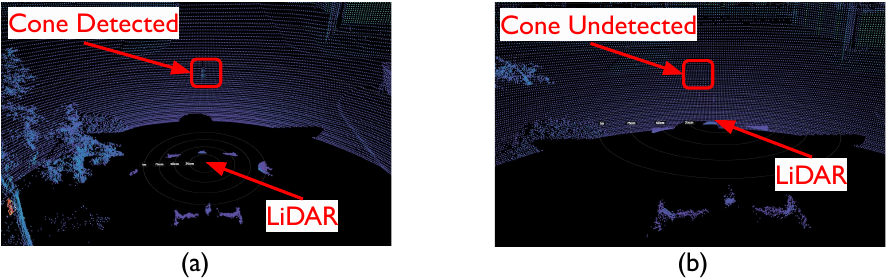}
\captionsetup{font=small}    
\caption{Effect of the ORA with a mirror at a 0° orientation. \textbf{(a)} Baseline LiDAR point cloud showing the traffic cone. \textbf{(b)} Under ORA, the mirror successfully removes the cone from perception.}
 \label{PCO: ORA}
\end{figure}

\noindent\textbf{Observation.}~A direct comparison of the LiDAR point clouds highlights the impact of the mirror on obstacle visibility. In the baseline runs (without a mirror), the traffic cone consistently produced a compact and dense cluster of points at its expected location, clearly indicating the presence of an object (see Figure~\ref{PCO: ORA}(a)). However, when a mirror was introduced, this cluster was entirely missing. Across all tested configurations, the cone was completely absent from the point cloud, creating the false impression of an unobstructed road (see Figure~\ref{PCO: ORA}~(b)). This demonstrates that even a mid-sized obstacle can be rendered invisible to a LiDAR sensor through passive optical redirection. A detailed analysis of how point cloud visibility varies with mirror angles is provided in Appendix~\ref{app:point_clouds_ORA}.

\begin{figure*}
\centering
\includegraphics[width=0.75\linewidth]{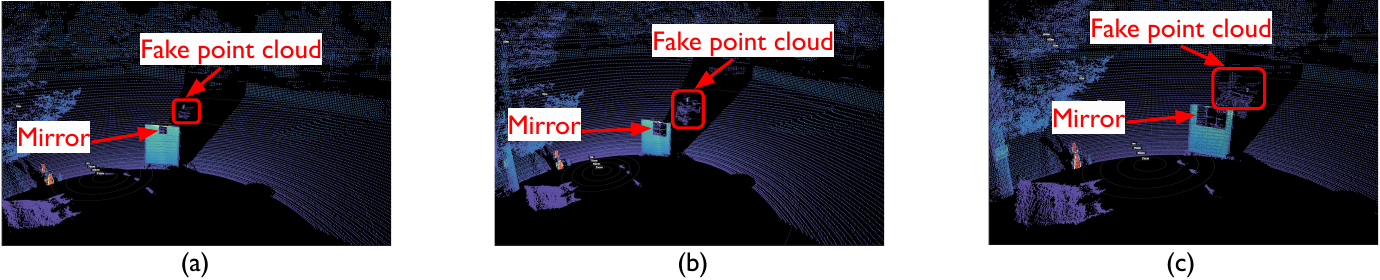}
\captionsetup{font=small}
    \caption{LiDAR point cloud results with different mirror configurations in OAA scenario. \textbf{(a)}: 2 mirror – a small cluster of reflected points is visible (circled) behind the mirror’s location. \textbf{(b)}: 4 mirrors – the cluster is larger and more clearly delineates a vertical surface. \textbf{(c)}: 6 mirrors – a high-density cluster appears, approximating the shape and size of a real vehicle.}
    \label{PC:Mirrors_size}
\end{figure*}

\subsection{Object Addition Attack Trials (OAA)}
The second set of experiments investigated the OAA scenario, which aims to generate \textit{fake LiDAR points} by reflecting environmental structures into the sensor’s field of view. These trials were conducted in the same open parking environment. At one edge of intersection, we set up a vertical white board on which we mounted a modular mirror array (see Figure~\ref{fig:Real_ORA_OAA}~(b)). The array consisted of 30\,cm$\times$30\,cm square mirror tiles; we could adjust how many tiles (surface area) and their arrangement.
We tested the mirror array at three tilt angles: 30$^\circ$, $45^\circ$, and 60$^\circ$ \,relative to the LiDAR’s forward-facing axis. These angles were chosen to span a range from moderate to steep reflection scenarios. For each angle, we also varied the \textit{reflective surface area} by configuring the array as: 2 tiles (0.18\,m$^2$ area), 4 tiles (2$\times$2 grid, 0.36\,m$^2$), 6 tiles (3$\times$2 grid, 0.6\,m$^2$).
We then conducted trials in which the vehicle approached the mirror installation from an initial distance of approximately 20\,m, driving at a steady speed of 8\,km/h. For safety reasons, the vehicle remained under manual control. We executed the intersection turn to maintain a full view of the potential fake point cluster. During each configuration, we continuously recorded full-resolution LiDAR scans, enabling analysis of how \textit{fake point clusters emerge, persist, or vanish} as a function of distance.

\noindent \textbf{Observation.}~The emergence of virtual points behind the mirror was consistently observed across all trials, confirming the mirror’s ability to inject false depth readings into the LiDAR data. The experimental trials revealed a direct and significant correlation between the mirror's surface area and the robustness of the generated virtual point cloud. As the reflective area was increased from 2 tiles (0.18\,m$^2$) to the largest six-tile array (0.6\,m$^2$), both the number of virtual points and their spatial density grew substantially. This progression is clearly illustrated in Figure~\ref{PC:Mirrors_size}, where the point cloud evolves from a sparse, fragmented cluster into a dense and cohesive structure that more convincingly resembles a solid object. This finding indicates that a larger mirror surface is considerably more effective at fabricating a believable virtual obstacle within the sensor data.
The mirror's tilt angle was also observed to be a critical factor, primarily influencing the location, timing, and persistence of the virtual point cloud during the vehicle's approach (see Figure \ref{fig:OAAappendix_angle}). At a shallow 30$^\circ$ \,angle, the geometric conditions for reflection were only met when the vehicle was very close to the mirror, causing the virtual points to appear abruptly at a short perceived distance and for a brief temporal window. Conversely, a steeper 60$^\circ$ \,angle created a stable reflection that was visible much earlier in the vehicle's approach, generating a persistent cluster of points at a greater perceived distance. The $45^\circ$ \,angle presented an intermediate case, producing a point cloud that was less stable than at 60$^\circ$ \,but persisted longer than at 30$^\circ$. This highlights a strategic trade-off between the perceived immediacy of the virtual obstacle and the temporal window of its appearance in the LiDAR data. A detailed geometric analysis of these angular effects is provided in Appendix~\ref{app:point_clouds_OAA}.

These preliminary results confirm that the properties of the virtual point cloud are fundamentally governed by the physical setup; the mirror's surface area, its tilt angle, and the dynamic distance between the sensor and the mirror.  For a predictive and generalizable understanding, it is essential to formally model this behavior. Therefore, the subsequent section develops an empirical model to characterize key artifact properties—such as the number of points, lateral offset, radial distance, and probability of appearance—as a function of the physical parameters of the attack setup.

\section{Empirical Modeling of Mirror-Induced LiDAR Artefacts}
\label{sec:empirical_modeling}
This section presents a comprehensive empirical characterization of the artifacts introduced during the OAA, using the data recorded from the experiments described previously. The modeling pipeline begins with the extraction of reflection-induced features from these point clouds, which are then used to derive an analytical model suitable for robust simulation and attack injection purposes.

\subsection{Artifact Segmentation and Feature Extraction}

From the outdoor experiments, we extracted instances of \textit{mirror-induced artifact points} by comparing mirror-present data to baseline data. Specifically, for each trial configuration, we performed \textit{frame differencing}: aligning frames from the mirror run with corresponding frames (same vehicle position) from the baseline run, and then identifying points that appear only when the mirror is present. This was facilitated by performing a fine-grained spatial registration (using ICP – Iterative Closest Point) between mirror and baseline point clouds to ensure any minor vehicle motion or tilt was compensated. The resulting differenced cloud highlights the “extra” points due to the mirror. These \textit{artifact clusters} (the fake objects) were clearly segmented behind the mirror position. We manually labeled these clusters in a subset of frames for ground truth and then applied radius-based clustering to automatically group points in remaining frames.

For each trial, we segmented the artifact clusters and annotated them with the instantaneous physical parameters of their creation: the LiDAR-to-mirror distance ($d$), mirror tilt angle ($\theta$), and mirror surface area ($A$). From each artifact cluster, we then extracted a set of quantitative features to serve as the dependent variables for our empirical models:

\noindent $\bullet$ \textbf{Artifact Location ($R_{\text{artifact}}$, $X_{\text{artifact}}$):}~We first compute the 3D centroid ($C_{\text{artifact}}$) of the point cluster. From this centroid, we derive its two primary geometric components for modeling: 
1) \textit{Radial Distance ($R_{\text{artifact}}$)} determine magnitude of the centroid vector, representing the artifact's apparent distance from the sensor.
2) \textit{Lateral Offset ($X_{\text{artifact}}$)} determine the displacement of the centroid from the LiDAR's central forward axis, representing its perceived sideways position.

\noindent $\bullet$  \textbf{Artifact Intensity ($N_{\text{artifact}}$):}~The total number of points in the artifact cluster per frame. This metric serves as a proxy for the artifact's salience or visibility to the perception system.

\noindent $\bullet$ \textbf{Artifact Appearance Probability ($P_{\text{app}}$):}~The empirical probability that an artifact cluster (with $N > 0$) appears in a LiDAR frame under a given configuration $(d, \theta, A)$.

This structured dataset, which maps the physical input parameters ($d, \theta, A$) to the observed artifact features ($R_{\text{artifact}}, X_{\text{artifact}}, N_{\text{artifact}}, P_{\text{app}}$), provides the empirical foundation for the predictive models developed in the next section.

\subsection{Empirical Model Formulation and Validation}

\begin{figure*}
\centering
    \includegraphics[width=\linewidth]{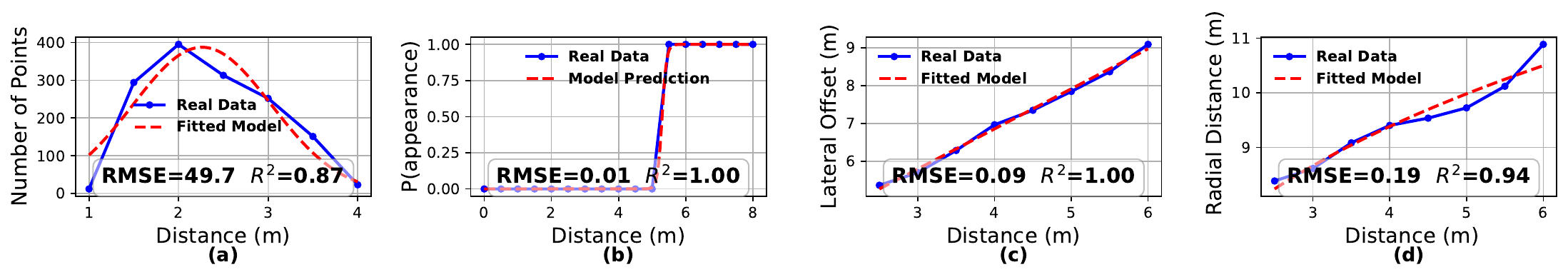}
    \captionsetup{font=small}
\caption{Comparison of model predictions (red dashed) with experimental data (blue solid) across representative configurations, illustrating the high fidelity of our empirical models. 
    \textbf{(a)} The Artifact Point Count model captures the bell-shaped intensity profile for a 30°, 0.36\,m$^2$ mirror. 
    \textbf{(b)} The Appearance Probability model delineates the detection window for a 60°, 0.6\,m$^2$ setup. 
    \textbf{(c, d)} The Lateral Offset and Radial Distance models accurately track the artifact location for a 45°, 0.36\,m$^2$ case.}
    \label{fig:model_validation_summary}
\end{figure*}

Using the extracted data, we derived analytical models that capture the dependency of mirror-induced artifacts on the geometric and reflective parameters ($d$, $\theta$, $A$). Each model is chosen to be physically interpretable while achieving a high goodness-of-fit on our data.

\noindent \tikz[baseline=(char.base)]\node[shape=circle, fill=black, inner sep=0.8pt, text=white] (char) {1}; \textbf{Artifact Location Model.}~We model the location of the artifact's centroid, $C_{\text{artifact}}$, by predicting its primary geometric components: its radial distance from the sensor ($R_{\text{artifact}}$) and its lateral offset from the LiDAR's central axis ($X_{\text{artifact}}$). This approach allows for a clean, interpretable model grounded in the principles of geometric optics and refined with empirical observations from our data.

\indent The artifact's location is a direct consequence of the two-hop path taken by the LiDAR pulse: from the sensor to the mirror, then from the mirror to a secondary surface, and back along the same path. The mirror's tilt angle, $\theta$, is the critical parameter, as it deflects the laser beam by an angle of approximately $2\theta$, which fundamentally determines the geometry of this two-hop path.

\noindent\textbf{Modeling the Artifact's Lateral Offset.}~The lateral offset, $X_{\text{artifact}}$, is the most direct consequence of the angular deflection. Based on simple trigonometry, the offset is proportional to the distance traveled and the tangent of the deflection angle. Our model captures this primary relationship:
\begin{equation}
    X_{\text{artifact}}(d, \theta) = c_X \cdot d \cdot \tan(2\theta) + \delta_X,
    \label{eq:lateral_offset_model}
\end{equation}
here, the term $d \cdot \tan(2\theta)$ represents the first-order geometric expectation. The fitted coefficient $c_X$ and offset $\delta_X$ are empirical parameters that account for the precise geometry of our experimental setup (e.g., the height of the LiDAR and mirror relative to the ground). This model accurately reflects our observation that the artifact is cast further to the side as both the distance $d$ and tilt angle $\theta$ increase.

\noindent\textbf{Modeling the Artifact's Radial Distance.}~The radial distance, $R_{\text{artifact}} = \|C_{\text{artifact}}\|$, corresponds to half the total travel distance of the laser pulse. This total path is necessarily longer than the direct round-trip to the mirror ($2d$) because of the secondary path leg created by the reflection.

While an ideal geometric model might suggest a simple trigonometric scaling (e.g., $R_{\text{artifact}} \propto d \cdot \sec(2\theta)$), we observed a key non-linear effect in our data: the growth of $R_{\text{artifact}}$ with distance $d$ was distinctly \textbf{sub-linear}. This phenomenon is caused by the LiDAR beam's natural divergence. As distance $d$ increases, the beam's footprint on the mirror grows. For a finite-sized mirror, an increasing fraction of the beam's energy ``spills" over the edges and does not contribute to the reflection. The points that are successfully reflected tend to originate from the parts of the beam that travel shorter paths, biasing the measured centroid and suppressing the apparent increase in distance.

To capture this physically-driven, sub-linear behavior, we model the radial distance using a power-law relationship, a standard method for representing such diminishing-return phenomena:
\begin{equation}\small
    R_{\text{artifact}}(d, \theta) = c_R \cdot d^{n_d} \cdot (1 + f(\theta)),
    \label{eq:radial_distance_model}
\end{equation}
where $c_R$ is a scaling coefficient, $d^{n_d}$ models the distance dependence. We empirically confirmed the exponent $n_d < 1$ (fitting to approximately $0.88$), which directly represents the sub-linear growth due to beam spillover. For the angular dependence, we found a second-order polynomial, $f(\theta) = a_1 \theta + a_2 \theta^2$, provides a robust fit. The fitted coefficients for our model are detailed in Appendix~\ref{app:model_params}.

 Together, Equations~\ref{eq:lateral_offset_model} and~\ref{eq:radial_distance_model} form our complete predictive model for the artifact's location. 
As shown in Figure~\ref{fig:model_validation_summary}~(c) and (d), this two-part location model achieves an excellent goodness-of-fit, with $R^2$ values of 1.00 and 0.94 respectively. 
Notably, the mirror's surface area, $A$, was found to have a negligible direct effect on the artifact's \textit{location}; as we will show next, its critical impact is on the artifact's \textit{intensity} and \textit{detection probability}.

\noindent \tikz[baseline=(char.base)]\node[shape=circle, fill=black, inner sep=0.8pt, text=white] (char) {2}; \textbf{Artifact Point Count Model.}~We next model the number of points, $N_{\text{artifact}}$, in the artifact cluster. This metric is a proxy for the artifact's salience and is governed by the amount of received optical power, $P_r$, that exceeds the LiDAR sensor's internal detection threshold, $P_{\text{th}}$.

Our theoretical starting point is the LiDAR range equation for a simple target. In this formulation, the received power is given by:
\begin{equation}\small
    P_r \propto \frac{P_t \cdot A_{\text{eff}}}{R^4},
    \label{eq:lidar_range_simple}
\end{equation}
where $P_t$ is the transmitted power, $A_{\text{eff}}$ is the target's effective cross-section, and $R$ is the total path distance. The strong $1/R^4$ decay arises from the two-way propagation of the signal.

However, this simple model is insufficient for our two-hop reflection scenario. The effective cross-section and the total path length are complex functions of our input parameters $(d, \theta, A)$. More importantly, our empirical data reveals a non-monotonic behavior: the number of artifact points first increases with distance, peaks at an optimal distance, and then falls off. This creates a distinct ``detection envelope". This is due to the interplay between the reflection geometry, which is unfavorable at very close ranges, and signal attenuation, which dominates at long ranges.

The probability of detecting a single point, $p_{\text{det}}$, is a function of the received power $P_r$ relative to the threshold $P_{\text{th}}$. This relationship results in the observed bell-shaped curve for detection probability as a function of distance. The Gaussian function is the canonical mathematical form for modeling such a distribution. We can therefore model the probability of detection at a distance $d$ (for a given angle $\theta$) as being proportional to a Gaussian distribution:
\begin{equation}\small
    p_{\text{det}}(d \mid \theta) \propto \exp\left(-\frac{(d - \mu(\theta))^2}{2\sigma^2}\right),
    \label{eq:gaussian_prob}
\end{equation}
where $\mu(\theta)$ is the optimal detection distance for a given angle, and $\sigma$ controls the width of the detection envelope.

The total number of artifact points, $N_{\text{artifact}}$, is proportional to the number of laser rays that strike the mirror multiplied by this detection probability. The number of rays is related to the mirror's effective area, which is influenced by its physical area $A$ and its projected area at angle $\theta$. Based on these physical considerations, we construct our final, comprehensive model by combining these factors:
\begin{equation}\small
    N_{\text{artifact}}(d, \theta, A) = c_0 \cdot A^{\beta} \cdot \cos^{\gamma}(\theta) \cdot \exp\left(-\frac{(d - \mu(\theta))^2}{2\sigma^2}\right),
    \label{eq:point_count_model_final}
\end{equation}

In this final model, the Gaussian term models the detection probability envelope as derived in Eq.~\ref{eq:gaussian_prob}. The term $A^{\beta}$ captures the sub-linear scaling with mirror area that we observed (fitted $\beta \approx 1.25$), and the $\cos^{\gamma}(\theta)$ term models the sharp angular falloff, with a fitted exponent $\gamma \approx$ 3.5 indicating significant signal loss beyond simple geometric projection. The model's ability to capture the intensity profile is shown in Figure~\ref{fig:model_validation_summary}~(a), achieving an $R^2$ of 0.87 for the selected experimental run.

\noindent \tikz[baseline=(char.base)]\node[shape=circle, fill=black, inner sep=0.8pt, text=white] (char) {3}; \textbf{Artifact Appearance Probability Model.}~Finally, we model the probability, $P_{\text{app}}$, that an artifact appears in the point cloud. This model must capture the boundary conditions where reflections are both geometrically possible and have sufficient signal strength to be registered by the sensor.

Our experimental data revealed that artifacts appear only within a specific ``appearance window" of distances. A key observation is that the start and end points of this window, $d_{\min}$ and $d_{\max}$, are not fixed; they shift significantly as a function of the mirror's tilt angle $\theta$ and area $A$. Standard monotonic or simple quadratic models proved insufficient as they cannot account for these dynamically shifting boundaries.
To accurately model this behavior, we construct a phenomenological model that explicitly represents a window with variable boundaries. The probability function is formulated as a product of two sigmoid functions, which creates a soft window with a rising edge at $d_{\min}$ and a falling edge at $d_{\max}$:
\begin{equation}\small
    P_{\text{app}}(d, \theta, A) = \frac{1}{1 + \exp(-k(d - d_{\min}))} \cdot \frac{1}{1 + \exp(k(d - d_{\max}))},
    \label{eq:prob_window}
\end{equation}
where $k$ is a parameter that controls the steepness of the window's edges. The boundaries themselves are modeled as linear functions of the physical parameters:
\begin{align}\small
    d_{\min}(\theta, A) &= b_0^{\min} + b_1 \theta + b_2 A, \label{eq:d_min} \\
    d_{\max}(\theta, A) &= b_0^{\max} + c_1 \theta + c_2 A, \label{eq:d_max}
\end{align}
where the empirically fitted coefficients $b_0^{\min}$, $b_1$, $b_2$, $b_0^{\max}$, $c_1$, and $c_2$ define the appearance window, which is linearly influenced by the mirror's angle ($\theta$) and area ($A$). All coefficients were empirically fitted to our experimental data, and their numerical values are provided in Appendix~\ref{app:model_params}.
Figure~\ref{fig:model_validation_summary}~(b) validates this approach, showing the model's ability to replicate the sharp on/off behavior of artifact appearance with R$^2$ of 1. 

\begin{figure}[t]
\centering
    \includegraphics[width=\linewidth]{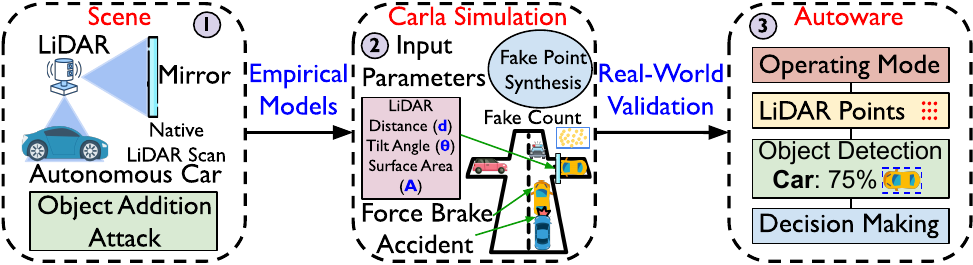}
\captionsetup{font=small}  
\caption{Overview of our end-to-end pipeline for modeling and evaluating mirror-based LiDAR attacks. \textbf{(1)} The process starts with a real-world Object Addition Attack scene, used to derive \textbf{Empirical Models}. \textbf{(2)} These models are integrated into the \textbf{CARLA simulator} to inject realistic fake points based on the vehicle’s state. \textbf{(3)} Key outcomes are validated through real-world testing on an \textbf{Autoware}-equipped vehicle, where the phantom object triggers a high-confidence false detection and a critical control failure.}
\label{fig:pipeline}
\end{figure}

With our models for artifact location, point count, and appearance probability, we establish a complete, parameterized description of mirror-induced LiDAR artifacts. The system-level evaluation follows a two-stage methodology, illustrated in Figure~\ref{fig:pipeline}. In the first stage, the empirical models drive a physics-informed attack injection pipeline implemented in the CARLA simulator. In the second stage, we validate the key behavioral failures observed in simulation through real-world experiments using an Autoware-equipped test vehicle. These two stages are detailed in the following sections.

\section{Simulation-Based Attack Injection in CARLA}
\label{sec:simulation}
After developing a set of high-fidelity empirical models, we integrate them into a simulation framework to evaluate the impact of mirror-based LiDAR attacks on autonomous vehicles.
Using the CARLA simulator \cite{dosovitskiy2017carla}, a cutting-edge open-source driving environment, we establish a \textit{physics-informed attack injection pipeline}. This framework provides a safe, repeatable, and scalable method for testing these attacks on a virtual AV with a standard autonomous driving agent. The simulation aims to first validate our models against real-world test results, and then to generalize the threat by exploring a wide range of attack parameters impossible to test physically.

\subsection{The Attack Injection Pipeline}

Our attack injection pipeline operates in real-time within the CARLA simulation loop. This process, formalized in Algorithm~\ref{alg:injection_pipeline}, consists of four distinct stages:

\noindent \tikz[baseline=(char.base)]\node[shape=circle, fill=black, inner sep=0.8pt, text=white] (char) {1}; \textbf{State Extraction.}~The pipeline begins by calculating the precise geometric relationship between the ego vehicle and the adversarial mirror. Using CARLA's APIs, we access the world-space transforms of both the LiDAR sensor and the mirror object. These are used to compute the three critical input parameters for our models: the instantaneous LiDAR-to-mirror distance ($d$), mirror tilt angle ($\theta$), and mirror surface area ($A$). This step effectively translates the dynamic 3D world of the simulator into the specific parametric inputs our empirical models require.

\begin{algorithm}[ht]\tiny
    \captionsetup{font=small}
    \caption{Model-Driven Artifact Injection Pipeline}
    \small
    \label{alg:injection_pipeline}
    \begin{algorithmic}[1]
        \Require Current mirror configuration $(d, \theta, A)$
        \Require Native LiDAR scan $\mathcal{L}_{\text{native}}$
        \Require Empirical models: $P_{\text{app}}, N_{\text{artifact}}, X_{\text{artifact}}, R_{\text{artifact}}$
        \State $p_{app} \leftarrow P_{\text{app}}(d, \theta, A)$
        \If{$\text{r}() < p_{app}$}
            \State $N \leftarrow \text{int}(N_{\text{artifact}}(d, \theta, A))$
            \State $R \leftarrow R_{\text{artifact}}(d, \theta, A)$
            \State $X \leftarrow X_{\text{artifact}}(d, \theta, A)$
            \State $C_{\text{artifact}} \leftarrow \text{ConvertTo3D}(R, X)$ \Comment{Convert to 3D point}
            \State $\mathcal{F} \leftarrow \emptyset$ \Comment{Initialize empty artifact cluster}
            \For{$i = 1$ to $N$}
                \State Sample point $\mathbf{p}_i \sim \mathcal{N}(C_{\text{artifact}}, \Sigma_{\text{fixed}})$
                \State Add $\mathbf{p}_i$ to $\mathcal{F}$
            \EndFor
            \State $\mathcal{L}_{\text{modified}} \leftarrow \mathcal{L}_{\text{native}} \cup \mathcal{F}$
        \Else
            \State $\mathcal{L}_{\text{modified}} \leftarrow \mathcal{L}_{\text{native}}$
        \EndIf
        \State \Return Modified LiDAR scan $\mathcal{L}_{\text{modified}}$
    \end{algorithmic}
\end{algorithm}

\noindent \tikz[baseline=(char.base)]\node[shape=circle, fill=black, inner sep=0.8pt, text=white] (char) {2}; \textbf{Probabilistic Attack Trigger.}~The extracted state vector is passed into a stochastic gate that determines whether an artifact should be injected. The inherent randomness in real-world sensor physics, where signals fluctuate near detection thresholds, is captured by our empirically derived ($P_{\text{app}}$). For a given state $(d, \theta, A)$, the model outputs the likelihood of artifact presence. A uniformly distributed random number $r \in [0, 1]$ is sampled, and the artifact is generated only if $r < P_{\text{app}}(d, \theta, A)$. The probabilistic mechanism reflects the intermittent flicker of marginal detections observed in practice.

\noindent \tikz[baseline=(char.base)]\node[shape=circle, fill=black, inner sep=0.8pt, text=white] (char) {3}; \textbf{Model-Driven Artifact Synthesis.}~Once triggered, the pipeline synthesizes a phantom object based entirely on the validated empirical models. All characteristics are derived from the same state vector $(d, \theta, A)$. The point count model $N_{\text{artifact}}$ determines the number of LiDAR points forming the artifact. Simultaneously, the location models $X_{\text{artifact}}$ and $R_{\text{artifact}}$ estimate the 3D centroid $C_{\text{artifact}}$. A cluster of $N$ points is then sampled from a multivariate Gaussian distribution $\mathcal{N}(C_{\text{artifact}}, \Sigma_{\text{fixed}})$, where $\Sigma_{\text{fixed}}$ is a small, constant covariance matrix. This process ensures a compact, spatially coherent artifact that resembles a physical object in the scene.

\noindent \tikz[baseline=(char.base)]\node[shape=circle, fill=black, inner sep=0.8pt, text=white] (char) {4}; \textbf{Point Cloud Injection.}~The final stage performs the insertion of the synthesized artifact cluster, $\mathcal{F}$, into the LiDAR data stream. The generated 3D points are merged with the native LiDAR scan, $\mathcal{L}{\text{native}}$, using a set union: $\mathcal{L}{\text{modified}} = \mathcal{L}_{\text{native}} \cup \mathcal{F}$. This modification is applied directly at the sensor-data level, prior to any downstream perception or control processing. As a result, the attack remains entirely transparent to the autonomy stack, which interprets the modified point cloud as authentic sensor input, enabling the observation of unfiltered system reactions.

\subsection{Scenario-Based Evaluation of AV Reactions}

The system-level impact of our attack was evaluated in a critical urban driving scenario designed in CARLA. This scenario enables observation of the complete event chain, from initial perception failure to a physical collision, thereby providing a clear measure of the attack's severity.

\noindent\textbf{Simulation Setup.}~Our primary scenario involves two vehicles approaching a T-intersection. The lead vehicle, which we term the ``Ego AV," is equipped with our full attack injection pipeline and a standard autonomous driving agent (\texttt{BehaviorAgent} in CARLA). A second ``Follower Vehicle" trails the Ego AV, governed by a basic car-following logic that maintains a safe distance based on its own perception. An adversarial mirror is placed at the corner of the T-intersection, positioned to trigger model-based artifact injection that simulates a phantom object in the Ego AV’s path.

\noindent\textbf{Attack Execution and Qualitative Outcome.}~As the Ego AV approaches the intersection, its position relative to the mirror continuously changes. Our injection pipeline evaluates the state $(d, \theta, A)$ in real-time. When the state enters a region with a high appearance probability, our framework injects a dense cluster of points into the Ego AV's LiDAR stream. This cluster is designed by our models to appear as a solid, crossing obstacle.

\begin{figure}[t]
    \centering
    \includegraphics[width=\linewidth]{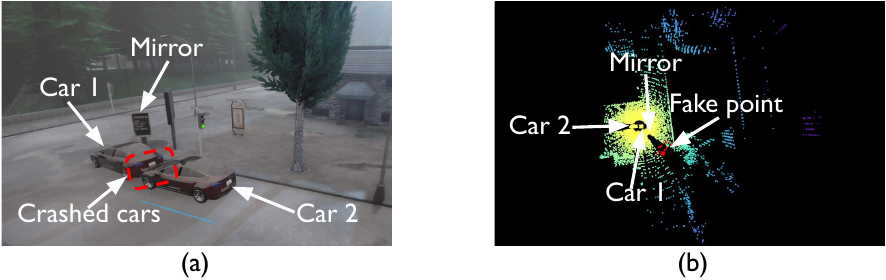}
    \caption{The catastrophic outcome of the OAA attack in simulation. The Ego AV (Car 1), having performed an emergency stop in response to a phantom obstacle, is struck from behind by a follower vehicle (Car 2). \textbf{(a)} The moment of collision in the CARLA environment. \textbf{(b)} The corresponding LiDAR view from Car 1, showing the injected ``Fake point" cluster that triggered the event.}
    \label{fig:crash_seq}
\end{figure}

\indent The AV's perception stack identifies the dense point cluster as a high-risk object. Interpreting it as a potential hazard entering the intersection, the safety-critical logic overrides the primary navigation objective and triggers an \textbf{emergency braking maneuver}. The Follower Vehicle, detecting only the sudden and unexpected deceleration of the lead car, fails to react and brake in time. This sequence consistently results in a high-speed rear-end collision, exposing a catastrophic failure mode induced by the passive optical attack. An example of this outcome is shown in Figure~\ref{fig:crash_seq}.

\begin{figure}[t]
    \centering
    \includegraphics[width=\linewidth]{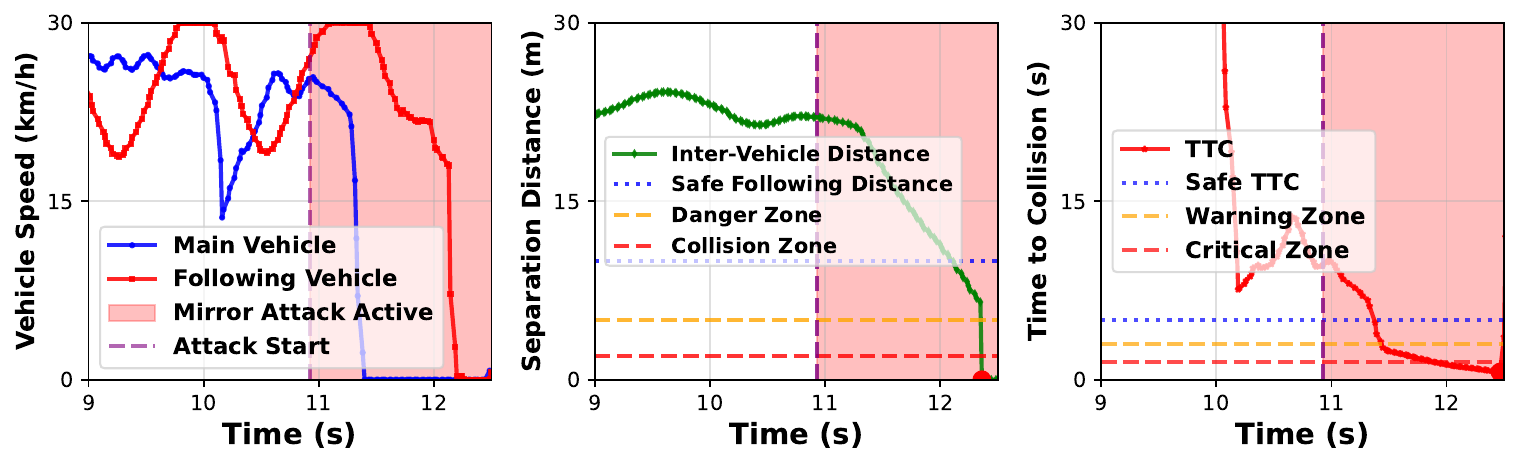}
    \caption{System-level failure analysis of a mirror-induced phantom obstacle attack. The plots show key dynamic metrics following the attack initiation at $t \approx 11.0$\,s. The Ego AV performs an emergency brake in response to the injected artifact, while the Follower Vehicle fails to react in time. Both the separation distance and TTC drop sharply, ultimately reaching zero, indicating a rear-end collision.}
    \label{fig:crash_calculation}
\end{figure}

\begin{figure*}[!ht]
\centering
    \includegraphics[width=0.8\linewidth]{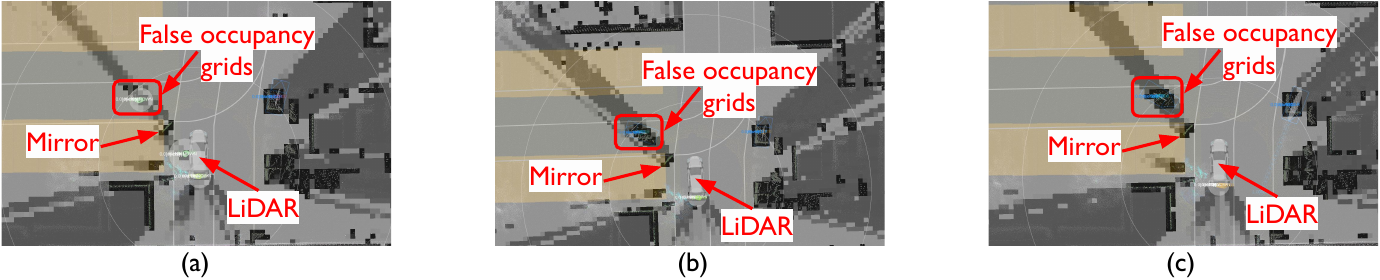}
    \captionsetup{font=small}
\caption{Occupancy grid maps under mirror-induced phantom object (OAA) conditions. \textbf{(a)}~2 mirrors – a thin line of occupied cells (black) appears just behind the mirror’s position. \textbf{(b)}~4 mirrors – a dense cluster of occupied cells forms, sufficient to trigger a false ``CAR" detection by the perception system. \textbf{(c)}~6 mirrors – a larger area is marked as occupied, closely matching the footprint of a real vehicle.}
    \label{Occupancy: OAA}
\end{figure*}

\noindent\textbf{Quantitative Impact Analysis.}~Figure~\ref{fig:crash_calculation} presents a time-aligned view of vehicle speed, inter-vehicle separation distance, and Time-to-Collision (TTC) during a representative mirror attack scenario.
The data reveal a clear causal chain initiated by the attack at $t \approx 11.0$\,s. Shortly after the artifact is injected, the Ego AV (blue) interprets it as a high-risk obstacle and initiates emergency braking, rapidly reducing its speed from over 25\,km/h to zero within approximately one second. The Follower Vehicle (red) continues at speed before reacting with a delay, causing the inter-vehicle gap to collapse. The separation distance not only drops below the safe following threshold but ultimately reaches zero, confirming a complete loss of spacing and physical contact.
In parallel, the TTC drops from a stable value to below the critical 3-second threshold and eventually reaches zero, indicating impact. The follower’s inability to respond in time confirms that the perception-layer deception triggers a cascading failure through the autonomy stack, culminating in a collision. This analysis demonstrates how a passive optical attack on LiDAR can compromise safety-critical behavior in autonomous vehicles.

\noindent\textbf{Effectiveness Across Configurations.}~The robustness of the attack was evaluated through a parameter sweep over diverse mirror configurations. Table~\ref{tab:attack-effectiveness} summarizes the outcomes for three representative setups, varying mirror distance, angle, and area. In all cases where a salient artifact was generated, the Ego AV initiated an emergency stop, resulting in a collision. These results confirm that the attack is not dependent on a finely tuned setup but remains effective across a range of physically plausible geometries. The results highlight the fragility of LiDAR-based perception to adversarial optics; the passive, physically-realizable attack consistently triggers a chain of misperception and unsafe vehicle behavior.

\begin{table}[t]
\centering
\captionsetup{font=small}
\caption{Attack Effectiveness Across Diverse Mirror Configurations}
\label{tab:attack-effectiveness}
\begin{tabular}{|c|c|c|c|}
\hline
\textbf{Mirror Config. ($d, \theta, A$)} & \textbf{Points Generated} & \textbf{AV Reaction} & \textbf{Collision?} \\
\hline
(4\,m, 30°, 0.18\,m$^2$) & $\approx$\,200 & Emergency Brake & Yes \\
(5\,m, 45°, 0.36\,m$^2$) & $\approx$\,55 & Emergency Brake & Yes \\
(7\,m, 60°, 0.60\,m$^2$) & $\approx$\,25& Emergency Brake & Yes \\
\hline
\end{tabular}
\end{table}

\section{Real-World Validation on an Autoware-Enabled Vehicle}
This section provides real-world validation of the mirror-based LiDAR attack by demonstrating its impact on a fully integrated autonomous driving system. Moving beyond simulation, we show how distortions introduced by mirrors affect the LiDAR perception pipeline in Autoware, leading to critical errors in object detection and triggering unsafe planning and control actions. The analysis highlights how a physically deployable optical setup can induce end-to-end failures in autonomous decision-making.

\subsection{Experimental Methodology}
The validation campaign utilized a Kia Soul EV equipped with an Ouster OS1-128 LiDAR and running the Autoware.AI software stack (as described in Section~\ref{Outdoor_exp}), with CenterPoint as the primary 3D object detector. CenterPoint was chosen due to its high accuracy and robustness in real-time LiDAR-based object detection, as demonstrated in both academic benchmarks and real-world autonomous driving systems \cite{yin2021centerpoint}. Two adversarial scenarios, OAA and ORA, were evaluated in an outdoor environment. The vehicle was operated in two distinct modes:

\noindent $\bullet$  \textbf{Manual Mode:}~A safety driver executed precise approach trajectories to log sensor data and the corresponding perception outputs without engaging vehicle control. This allowed for a clean analysis of how the perception system interprets adversarial inputs.

\noindent $\bullet$  \textbf{Autonomous Mode:}~The vehicle's full autonomy stack was engaged, making it responsible for perception, planning, and control. This mode was used to observe the system's unscripted, emergent behaviors in response to the attacks.

\indent Each scenario was tested across a parameter grid of three mirror surface areas (0.18, 0.36, and 0.6\,m$^2$) and three tilt angles (30$^\circ$, 45$^\circ$, and 60$^\circ$). For each run, we recorded synchronized logs of the occupancy grid maps, detected object lists, and the final vehicle control signals (throttle, brake, steering).

\subsection{Perception-Level Impact}

\noindent \tikz[baseline=(char.base)]\node[shape=circle,draw,inner sep=0.8pt] (char) {1}; \textbf{OAA: Phantom Object Generation and Misclassification.}~Under the OAA, maliciously reflected LiDAR returns corrupt the vehicle's perception, leading to cascading failures in higher-level modules. The first significant consequence is observed in the occupancy grid. As shown in Figure~\ref{Occupancy: OAA}, even with a small mirror composed of two panels, the mapping module marks several cells as ``occupied". As the mirror area increases to four and six panels, the number of occupied cells grows significantly, forming a dense, vehicle-sized region. The mapping module integrates these false returns over multiple frames, cementing the phantom object's existence within the vehicle's environmental model.

\begin{figure*}[!ht]
\centering
    \includegraphics[width=0.8\linewidth]{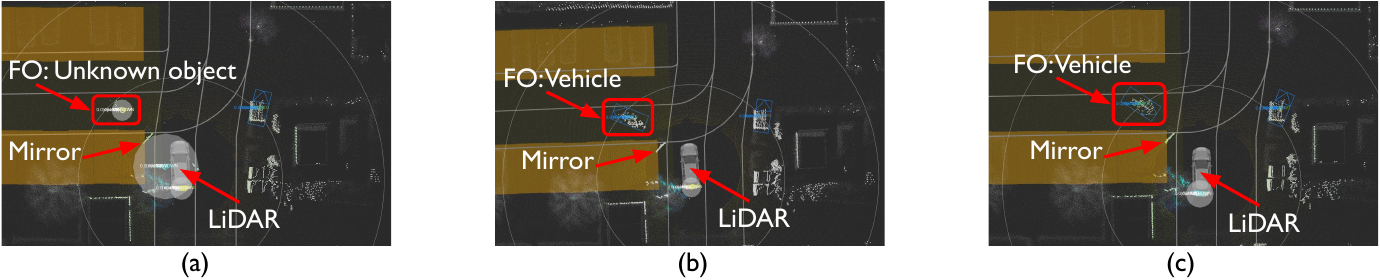}
    \captionsetup{font=small}
\caption{Autoware 3D object detection outcomes under mirror-induced phantom conditions. LiDAR point clouds are visualized with the corresponding detector bounding boxes. \textbf{(a)}~2 mirrors – the detector identifies an “UNKNOWN” object (white circle) with low confidence at the location of the fake points. \textbf{(b)}~4 mirrors – the cluster is labeled as a Vehicle (blue box) with approximately 65\% confidence. \textbf{(c)}~6 mirrors – the detection confidence increases to 74\%, and the phantom is firmly classified as a Vehicle.}
    \label{FO:object detection}
\end{figure*}

\indent The corruption extends beyond geometry to the semantic level. Autoware’s object detector interprets the dense cluster of phantom points as a valid obstacle, resulting in erroneous classifications. As shown in Figure~\ref{FO:object detection}, a small mirror array produces a low-confidence detection labeled as ``UNKNOWN", which, despite the uncertainty, is still treated as an obstacle by downstream modules. As the mirror area and point density increase, the system’s interpretation intensifies: with four mirrors, the artifact is misclassified as a ``CAR" with 65\% confidence, rising to 74\% with six mirrors. The detector effectively hallucinates a plausible, high-confidence object where none exists, revealing a critical breakdown in perception integrity.

\begin{figure}[t]
\centering
    \includegraphics[width=\linewidth]{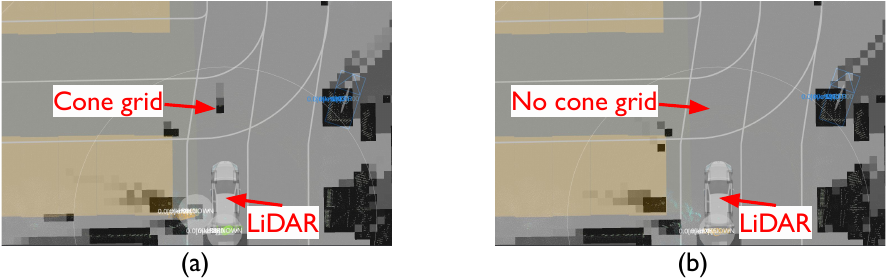}
    \captionsetup{font=small}
\caption{Occupancy grid comparison for ORA. \textbf{(a)}~Baseline without mirror – the cone’s position is marked as occupied (black cell), correctly indicating the presence of an obstacle. \textbf{(b)}~With mirror (ORA active) – the same region is falsely marked as free space (no occupied cell), as the LiDAR fails to receive returns from the cone.}
 \label{Ocuppancy: ORA and without ORA}
\end{figure}

\noindent \tikz[baseline=(char.base)]\node[shape=circle,draw,inner sep=0.8pt] (char) {2}; \textbf{ORA: Silent Obstacle Removal.}~The ORA scenario introduces a particularly deceptive failure mode by actively replacing legitimate obstacle data with plausible environmental signals. In this setup, mirrors are positioned to deflect LiDAR beams that would otherwise strike a real obstacle (e.g., a traffic cone) and redirect them toward the ground plane. As a result, the perception pipeline receives structurally valid returns, which Autoware’s ground segmentation algorithm interprets as a continuous, traversable surface. Figure~\ref{Ocuppancy: ORA and without ORA} illustrates this effect in the occupancy grid. In the baseline case without a mirror, the cone is correctly marked as an obstacle (black cell). When the mirror is introduced, the same region is falsely marked as free space, no occupied cells appear, despite the physical presence of the cone and mirror. This “data replacement” effect creates a high-confidence illusion of a clear path and effectively erases the real-world hazard from the vehicle’s environmental model. We found this removal effect to be robust across all tested mirror angles (see Figure~\ref{PC:OORA_angles-appendix}). This active deception is more likely to evade anomaly detection systems, which often search for sensor voids or dropouts, and directly compromises any safety architecture that trusts LiDAR as a ground-truth sensor for obstacle detection.

\subsection{System-Level Behavior and AV Decisions}
Having shown how the OAA scenario corrupts the perception stack, we now examine its impact on the vehicle’s real-time decision-making and control. The following experiments were conducted in \textbf{autonomous mode}, with Autoware’s planning and control modules fully active. We focus on the OAA scenario to illustrate how a phantom object can trigger hazardous high-level driving behaviors.

\noindent \tikz[baseline=(char.base)]\node[shape=circle, fill=black, inner sep=0.8pt, text=white] (char) {1}; \textit{Experimental Context and Methodology.}~The tests were conducted using the most effective OAA configuration, employing a 6-mirror array to generate a high-density phantom object. The vehicle was instructed to follow a route leading toward the mirror setup, simulating a common use case such as turning at an intersection. The mirror tilt angle was systematically varied across 30$^\circ$, 45$^\circ$, and 60$^\circ$, enabling analysis of how the attack geometry influences the timing and severity of the vehicle’s reaction. All trials were performed in fully autonomous mode, with a safety driver present solely for emergency intervention. The vehicle maintained a speed of 8\,km/h throughout the runs, aligning with the parking lot speed limit and ensuring safety during testing.

\begin{figure*}[!ht]
\centering
\includegraphics[width=0.7\linewidth]{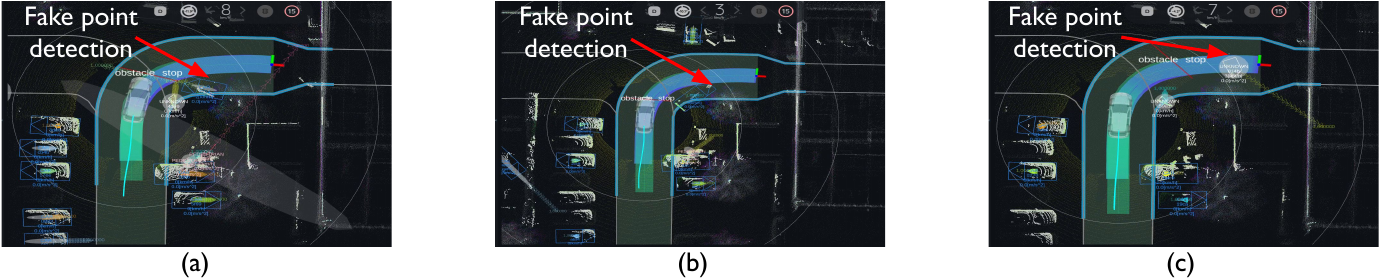}
\captionsetup{font=small}
\caption{Autoware’s RViz view showing phantom detections in the 6-mirror scenario at different mirror angles.
\textbf{(a)} 30$^\circ$: A phantom vehicle (blue ``Vehicle'' box) appears near the car, triggering an immediate stop
(\texttt{obstacle\_stop}). \textbf{(b)} 45$^\circ$: The phantom appears earlier, causing brief deceleration before stopping.
\textbf{(c)} 60$^\circ$: An unknown object (white ``UNKNOWN'' box) is detected ahead, leading to a full emergency stop.}
\label{fig:autonomos}
\end{figure*}

\noindent \tikz[baseline=(char.base)]\node[shape=circle, fill=black, inner sep=0.8pt, text=white] (char) {2}; \textit{Observed System Responses to Phantom Obstacles.}~Our analysis of the synchronized logs, including the RViz visualizations shown in Figure~\ref{fig:autonomos}, revealed that the vehicle's planner consistently interpreted the phantom as a real threat. In every successful attack, the system engaged its \textit{obstacle stop} procedure, but the nature of this response varied significantly with the attack's geometry, leading to distinct and dangerous failure modes.

\noindent $\bullet$ \textbf{Late-Onset Detection and High-Jerk Braking (30$^\circ$ Angle):}~At a shallow 30$^\circ$ angle, the reflective geometry only produced a detectable phantom when the vehicle was within a few meters of the mirrors. As shown in Figure \ref{fig:autonomos}~(a), the vehicle has almost completed the turn's apex when the phantom object (blue box) appears abruptly at extremely close range. This late-onset detection created a critically low TTC, forcing the motion planner to issue an immediate emergency \textit{obstacle stop} command. The vehicle halted with an extremely high jerk profile, a maneuver that, while logically correct from the system's perspective, would be highly dangerous in a real traffic environment and would likely cause a rear-end collision by a following vehicle.

\noindent \textbullet{} \textbf{Intermediate Detection and Mid-Turn Stop (45$^\circ$ Angle):}~
Increasing the angle to 45$^\circ$ widens the window for reflection. Consequently, the phantom object
is detected earlier in the vehicle's approach, as it begins to enter the curve but before it reaches the apex.
The RViz view shows (see Figure \ref{fig:autonomos}~(b)) the vehicle braking while still on the initial part
of the turn. The planner has slightly more time to react compared to the 30$^\circ$ case, but the outcome is
still a full stop in the middle of a dynamic maneuver. This scenario can be particularly hazardous if it leads
to detection flickering, causing decision instability (oscillating between braking and proceeding) and creating
an unpredictable hazard for other road users.

\noindent $\bullet$ \textbf{Early Detection and Pre-Turn Path Obstruction (60$^\circ$ Angle):}~At 60$^\circ$, the mirror presents a large, effective reflective surface to the approaching vehicle. This results in the earliest possible detection, occurring well before the vehicle even enters the turn. The image clearly shows the phantom object appearing as a full obstruction far down the planned trajectory (the light white circle). The system, identifying a high-confidence obstacle directly in its path, immediately terminates the plan and commands a full stop. While an early stop might seem safer, it is still a hazardous phantom braking event. The vehicle stops for no apparent reason long before an intersection, disrupting traffic flow and creating a dangerously unpredictable situation for human drivers who cannot see the phantom obstacle.

\noindent \tikz[baseline=(char.base)]\node[shape=circle, fill=black, inner sep=0.8pt, text=white] (char) {3}; \textit{Broader Safety and Traffic Implications.}~This analysis demonstrates that the OAA can trigger a spectrum of dangerous behaviors, all of which constitute \textit{errors of commission}. The vehicle is induced to perform an unnecessary and hazardous action, from a sudden, high-g emergency stop to an inexplicable halt in a live traffic lane. The severity and timing of this failure are directly tunable by the attacker's choice of mirror angle.
Although the tests were conducted at a low speed of 8\,km/h to ensure safety during repeated trials, autonomous vehicles typically approach intersections at higher speeds, often between 10 and 20\,km/h. At such speeds, the risks become significantly greater, as reduced reaction time and increased momentum can lead to more dangerous outcomes, including rear-end collisions or loss of control during abrupt braking. 
Crucially, in every case, the Autoware stack performed exactly as it was designed to, executing a logical, defensive maneuver based on the sensory information it received. The vulnerability lies not in the planning software but at the physical-sensor interface, where the vehicle's perception of reality was successfully manipulated. This proves that robust autonomous vehicle safety must extend beyond algorithmic integrity to include foundational defenses against the physical-world deception of its primary sensors.

\section{Discussion and Countermeasures}

Our findings show that mirror-based attacks exploit core assumptions in LiDAR perception, creating serious safety risks for autonomous systems. This section explores potential defense strategies, evaluates three implemented countermeasures, discusses study limitations, and outlines directions for future research.

\subsection{Defense Strategies}

Countermeasures against mirror-based LiDAR attacks fall into hardware-based and software-based approaches. We explore three strategies: thermal camera augmentation for cross-modal verification, multi-sensor fusion to improve robustness, and Light Fingerprinting to detect reflective surfaces.

\noindent \tikz[baseline=(char.base)]\node[shape=circle,draw,inner sep=0.8pt] (char) {1}; \textbf{Thermal Camera Augmentation.}~The fundamental principle of this defense is to leverage a different sensing modality that is not susceptible to the same physical deception. Thermal (long-wave infrared) cameras detect emitted heat rather than reflected light. In OAA, the phantom object is a LiDAR-specific illusion that lacks physical substance and therefore has no corresponding heat source. A thermal camera observing the phantom's perceived location would see only the ambient background temperature, directly falsifying the LiDAR detection. In ORA, the mirror itself becomes the detectable anomaly. Its surface typically presents a thermal signature distinct from the surrounding road, creating a clear thermal discontinuity that reveals the presence of the adversarial element. In both cases, the thermal modality provides critical information to either invalidate a phantom or detect the tool of deception.

\begin{figure}[t]
\centering
\includegraphics[width=\linewidth]{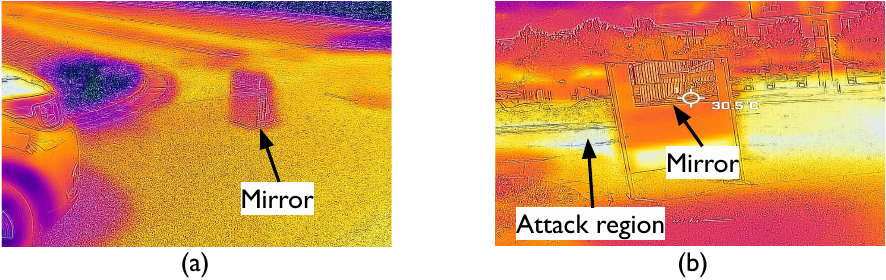}
\captionsetup{font=small}
\caption{Thermal imagery of mirror attack scenarios. \textbf{(a)}~ORA: The mirror is visible as a cool (purple) surface against the warmer (yellow) road. \textbf{(b)}~OAA: The phantom object region shows no thermal anomaly, matching the ground temperature and confirming it is not a real object.}
 \label{Thermal}
\end{figure}

\noindent \textbf{Implementation and Evaluation.}~We integrated a FLIR One thermal camera into our test vehicle and developed a detection module to identify anomalous thermal regions. As shown in Figure~\ref{Thermal}, the camera proved effective in both scenarios. In the ORA case (\textbf{a}), the mirror panel is clearly delineated as a cool (purple) rectangular surface against the warmer (yellow) asphalt, creating a distinct thermal boundary that allows for direct detection of the adversarial element. In the OAA scenario (\textbf{b}), the thermal image shows that the region where LiDAR detected a phantom vehicle is thermally uniform with its surroundings. The absence of any localized heat source different from the ground plane confirms the object's non-physical nature.

\noindent \textbf{Limitations.}~Despite its promise, thermal imaging is not a panacea. Its efficacy can be reduced by environmental factors, such as high ambient temperatures that minimize the thermal contrast between objects (thermal crossover). The resolution of thermal cameras is typically lower than that of RGB cameras, making it difficult to detect small or distant mirrors. Furthermore, practical deployment involves challenges such as cost, calibration, and maintenance, as lenses can be obscured by rain or dirt.

\noindent \tikz[baseline=(char.base)]\node[shape=circle,draw,inner sep=0.8pt] (char) {2}; \textbf{Multi-Sensor Fusion.}~Another potential defense strategy involves leveraging the inherent redundancy of an autonomous vehicle's multi-modal sensor suite. Standard sensor fusion algorithms typically operate on the principle of data reinforcement, assuming that different sensors will provide consistent, corroborating evidence about the state of the world. Mirror-based attacks are designed to violate this assumption by creating a fundamental \textit{disagreement} between modalities. An OAA creates a phantom visible only to LiDAR, while an ORA can make a real object invisible only to LiDAR. A defense can therefore be built around a ``fusion consistency monitor" designed to detect and flag these specific inter-sensor conflicts as anomalies.

\noindent \textbf{Conceptual Implementation and Mitigation.}~This defense would be implemented as a logic layer that continuously cross-verifies detections between LiDAR and other sensors, such as RGB cameras or radar. Aligned with proposals such as that of \cite{liu2021seeing}, the system would monitor for high-confidence inconsistencies. For instance, in an OAA scenario, a LiDAR-detected object that lacks a corresponding visual object in the camera feed would be flagged as anomalous. Conversely, in an ORA scenario, a visually-detected obstacle that corresponds to a region of navigable free space in the LiDAR data would trigger a similar alert. Upon detecting such a conflict, the system could respond by significantly lowering the confidence score of the suspect perception, or by triggering a safe-state behavior, such as reducing speed, until the conflict is resolved by subsequent sensor readings.

\noindent \textbf{Limitations and Challenges.}~The primary challenge of this approach lies in robustly disambiguating malicious inconsistencies from benign, modality-specific sensor failures. A naively implemented consistency check could generate high-risk false alarms. For example, in low-light conditions, a camera will naturally fail to detect a real obstacle that is clearly visible to LiDAR; a simple check would incorrectly discard this valid detection as a phantom. Similarly, benign physical phenomena, such as a retro-reflective road sign being highly salient to a camera but producing sparse returns for LiDAR, can also create disagreement. Therefore, developing a sophisticated consistency model that accounts for varying environmental conditions and the unique operational characteristics of each sensor without introducing an unacceptable rate of false positives is a non-trivial research challenge. Other advanced concepts, such as using redundant LiDARs at different physical locations to break the attack's geometric constraints, also warrant future exploration.

\noindent \tikz[baseline=(char.base)]\node[shape=circle,draw,inner sep=0.8pt] (char) {3}; \textbf{The Light Fingerprinting Framework.}~A more fundamental, software-based defense can be developed by analyzing the physical signature of the LiDAR returns themselves. The vulnerability we exploit stems from perception systems treating all LiDAR points as geometrically equal. This prospective defense would differentiate points based on their ``optical fingerprint," using attributes available in modern LiDAR sensors such as intensity, calibrated reflectivity, and pulse width. The core physical insight is that the reflection of a laser pulse differs fundamentally based on the target's material properties. Most real-world surfaces are \textit{diffuse reflectors} that scatter light, while mirrors are \textit{specular reflectors} that do not. A return pulse that is ``too perfect" is therefore a strong physical indicator of a specular, mirror-like reflection.

\noindent \textbf{Conceptual Implementation and Mitigation.}~The conceptual implementation involves training a deep neural network to classify the optical fingerprint of each LiDAR point in real-time. This would produce a semantically enriched point cloud where points are tagged with material properties, most critically `specular-reflective`. This material-aware approach could directly counter mirror-based attacks by providing the perception system with crucial context. For instance, the system could learn to treat objects composed of `specular-reflective` points as non-physical artifacts (mitigating OAA) or to identify a mirror placed on the road as a physical anomaly (mitigating ORA). By reasoning about the material properties of detected surfaces, the system can better distinguish real-world hazards from sensor-level deceptions.

\noindent \textbf{Challenges and Future Work.}~While conceptually powerful, this approach presents a clear agenda for future research. The primary hurdles are the creation of large-scale, material-annotated LiDAR datasets required for training robust classifiers and the development of computationally efficient models for real-time analysis. Overcoming these challenges is a critical next step toward building LiDAR perception systems that are resilient to a wider class of physical-layer attacks. 

\section{Limitations and Future Work}
\noindent \textbf{Generality Across LiDAR Technologies.}~Our experiments utilized a specific family of spinning LiDARs. While the underlying time-of-flight principle is common, future work should validate these findings across a wider range of sensor technologies, including FMCW LiDAR and systems that support full-waveform analysis.

\noindent \textbf{Influence of Environmental Conditions.}~Our tests were conducted exclusively in clear, dry weather. The influence of adverse conditions like rain or fog, which could unpredictably alter mirror reflectivity and LiDAR beam propagation, remains an important area for systematic investigation, both in simulation and controlled real-world tests.

\noindent \textbf{High-Speed Vehicle Dynamics.}~For safety reasons, our real-world experiments were limited to low-speed maneuvers. Evaluating the risks at higher speeds, including the potential for phantom-braking-induced collisions in highway scenarios, remains a critical direction for future on-road testing in controlled environments.


\noindent \textbf{Integration and Validation of Defenses.}~While this paper proposed several countermeasures, their full integration and validation within an operational AV stack  were beyond our initial scope. Future work should focus on this deep integration, including implementing the thermal-camera-based verification and developing the proposed software-based anomaly detectors. 

\section{Related Work}
\label{related work}
Our research intersects several key areas:  LiDAR spoofBelow, we summarize the most relevant prior work and highlighthow our contributions advance the current state of the art.

\noindent\textbf{LiDAR Spoofing and Reflective Artifacts.}~Most known LiDAR spoofing attacks are active, involving external lasers that inject fake pulses into the sensor to create phantom objects at arbitrary distances \cite{petit2015remote, cao2019adversarial, fukunagarandom,hau2021object}. These attacks are effective but require specialized hardware and line-of-sight access to the target sensor.
Reflective artifacts caused by surfaces like wet roads, glass, or metal have been observed in real environments \cite{Yang2011, damodaran2023experimental, tibebu2021lidar, yang2024evaluating, piroli2023towards}, but are typically treated as incidental measurement noise rather than deliberate threats. Our work is the first to frame mirror-based reflections as a passive, controllable, and repeatable attack vector capable of both object removal and object addition.

\noindent\textbf{Physical Adversarial Attacks.}~Several studies have explored physical-world attacks that reduce LiDAR performance through signal attenuation. For example, obscurants such as fog, smoke, or aerosols can scatter laser pulses, causing reduced range or blind spots \cite{rivero2021effect, dong2023benchmarking, piroli2023towards}. 
Similarly, other physical removal attacks can cause objects to disappear from perception, for instance by using reflective materials to create data voids \cite{kobayashiinvisible} or by exploiting sensor firmware limitations \cite{cao2023you}. 
These techniques remove information but do not create new, coherent structures. In contrast, our mirror-based approach manipulates the beam path, allowing an attacker to both hide real objects and fabricate plausible fake ones with spatial consistency.

\noindent\textbf{Defenses and Sensor‑Fusion Robustness.}~Defenses against LiDAR spoofing often rely on statistical filtering, temporal consistency checks, or signal-level validation to identify abnormal returns \cite{sun2020towards,hau2021shadow}. More advanced approaches use multi-sensor fusion to cross-validate LiDAR detections with camera or radar data, flagging inconsistencies such as a LiDAR object with no visual counterpart \cite{cao2021invisible, liu2021seeing,hallyburton2022security}.
These methods assume that cross-modal inconsistencies are rare and indicative of spoofing. However, real-world conditions often cause benign mismatches, such as camera failure in low light or occlusion. Our mirror-based attack exploits this ambiguity by generating a plausible LiDAR detection without a visual counterpart. The resulting inconsistency resembles a legitimate sensor mismatch, making it difficult for fusion algorithms to dismiss. As a result, the AV may default to conservative actions like emergency braking, even when no real threat exists.
\section{Conclusion}

This work presents the first comprehensive analysis of mirror-based LiDAR deception attacks on autonomous vehicles. We show that by exploiting the physics of specular reflection, an adversary can inject phantom obstacles or erase real ones using only inexpensive mirrors. Experiments on a full AV platform, with commercial-grade LiDAR and the Autoware stack, demonstrate that these are practical threats capable of triggering critical safety failures, such as abrupt emergency braking and failure to yield.
Our contributions include a foundational methodology and dataset to support the development of \textit{reflection-aware perception systems}. By integrating our models into a CARLA-based simulation framework, we also enable safe, scalable, and repeatable evaluation of this emerging threat.

\bibliographystyle{IEEEtran}
\bibliography{sample.bib}{}

\begin{thebibliography}{10}
\providecommand{\url}[1]{#1}
\csname url@samestyle\endcsname
\providecommand{\newblock}{\relax}
\providecommand{\bibinfo}[2]{#2}
\providecommand{\BIBentrySTDinterwordspacing}{\spaceskip=0pt\relax}
\providecommand{\BIBentryALTinterwordstretchfactor}{4}
\providecommand{\BIBentryALTinterwordspacing}{\spaceskip=\fontdimen2\font plus
\BIBentryALTinterwordstretchfactor\fontdimen3\font minus \fontdimen4\font\relax}
\providecommand{\BIBforeignlanguage}[2]{{%
\expandafter\ifx\csname l@#1\endcsname\relax
\typeout{** WARNING: IEEEtran.bst: No hyphenation pattern has been}%
\typeout{** loaded for the language `#1'. Using the pattern for}%
\typeout{** the default language instead.}%
\else
\language=\csname l@#1\endcsname
\fi
#2}}
\providecommand{\BIBdecl}{\relax}
\BIBdecl

\bibitem{zhang2024lidar}
Y.~Zhang, P.~Shi, and J.~Li, ``Lidar-based place recognition for autonomous driving: A survey,'' \emph{ACM Computing Surveys}, vol.~57, no.~4, pp. 1--36, 2024.

\bibitem{leong2024lidar}
P.~Y. Leong and N.~S. Ahmad, ``Lidar-based obstacle avoidance with autonomous vehicles: A comprehensive review,'' \emph{IEEE Access}, 2024.

\bibitem{bhupathiraju2023emi}
S.~H.~V. Bhupathiraju, J.~Sheldon, L.~A. Bauer, V.~Bindschaedler, T.~Sugawara, and S.~Rampazzi, ``{EMI-LiDAR}: Uncovering vulnerabilities of lidar sensors in autonomous driving setting using electromagnetic interference,'' in \emph{Proceedings of the 16th ACM Conference on Security and Privacy in Wireless and Mobile Networks}, 2023, pp. 329--340.

\bibitem{li2024detection}
Y.~Li, X.~Zhao, and S.~Schwertfeger, ``Detection and utilization of reflections in lidar scans through plane optimization and plane slam,'' \emph{Sensors}, vol.~24, no.~15, p. 4794, 2024.

\bibitem{suzuki2024wip}
R.~Suzuki, T.~Sato, Y.~Hayakawa, K.~Ikeda, O.~Sako, R.~Nagata, Q.~A. Chen, and K.~Yoshioka, ``Wip: Towards practical lidar spoofing attack against vehicles driving at cruising speeds,'' in \emph{ISOC Symposium on Vehicle Security and Privacy (VehicleSec)}, 2024.

\bibitem{Yang2011}
S.-W. Yang and C.-C. Wang, ``{On Solving Mirror Reflection in LIDAR Sensing},'' \emph{IEEE/ASME Transactions on Mechatronics}, vol.~16, no.~2, pp. 255--265, 2011.

\bibitem{Zhao2020}
X.~Zhao, Z.~Yang, and S.~Schwertfeger, ``{Mapping with Reflection -- Detection and Utilization of Reflection in 3D Lidar Scans},'' in \emph{Proceedings of the IEEE International Symposium on Safety, Security, and Rescue Robotics (SSRR)}, Abu Dhabi, UAE, 2020, pp. 27--33.

\bibitem{petit2015remote}
J.~Petit, B.~Stottelaar, M.~Feiri, and F.~Kargl, ``Remote attacks on automated vehicles sensors: Experiments on camera and lidar,'' \emph{Black Hat Europe}, vol.~11, no. 2015, p. 995, 2015.

\bibitem{sato2023lidar}
T.~Sato, Y.~Hayakawa, R.~Suzuki, Y.~Shiiki, K.~Yoshioka, and Q.~A. Chen, ``{LiDAR} spoofing meets the new-gen: Capability improvements, broken assumptions, and new attack strategies,'' \emph{arXiv preprint arXiv:2303.10555}, 2023.

\bibitem{jin2023pla}
Z.~Jin, X.~Ji, Y.~Cheng, B.~Yang, C.~Yan, and W.~Xu, ``{Pla-lidar}: Physical laser attacks against lidar-based {3d} object detection in autonomous vehicle,'' in \emph{2023 IEEE Symposium on Security and Privacy (SP)}.\hskip 1em plus 0.5em minus 0.4em\relax IEEE, 2023, pp. 1822--1839.

\bibitem{shin2017illusion}
H.~Shin, D.~Kim, Y.~Kwon, and Y.~Kim, ``Illusion and dazzle: Adversarial optical channel exploits against lidars for automotive applications,'' in \emph{Cryptographic Hardware and Embedded Systems--CHES 2017: 19th International Conference, Taipei, Taiwan, September 25-28, 2017, Proceedings}.\hskip 1em plus 0.5em minus 0.4em\relax Springer, 2017, pp. 445--467.

\bibitem{wang2023adversarial}
J.~Wang, F.~Li, X.~Zhang, and H.~Sun, ``Adversarial obstacle generation against lidar-based {3d} object detection,'' \emph{IEEE Transactions on Multimedia}, 2023.

\bibitem{liu2021seeing}
J.~Liu and J.-M. Park, ``“seeing is not always believing”: detecting perception error attacks against autonomous vehicles,'' \emph{IEEE Transactions on Dependable and Secure Computing}, vol.~18, no.~5, pp. 2209--2223, 2021.

\bibitem{yan2016can}
C.~Yan, W.~Xu, and J.~Liu, ``Can you trust autonomous vehicles: Contactless attacks against sensors of self-driving vehicle,'' \emph{Def Con}, vol.~24, no.~8, p. 109, 2016.

\bibitem{tu2020physically}
J.~Tu, M.~Ren, S.~Manivasagam, M.~Liang, B.~Yang, R.~Du, F.~Cheng, and R.~Urtasun, ``Physically realizable adversarial examples for lidar object detection,'' in \emph{Proceedings of the IEEE/CVF conference on computer vision and pattern recognition}, 2020, pp. 13\,716--13\,725.

\bibitem{cao2021invisible}
Y.~Cao, N.~Wang, C.~Xiao, D.~Yang, J.~Fang, R.~Yang, Q.~A. Chen, M.~Liu, and B.~Li, ``Invisible for both camera and lidar: Security of multi-sensor fusion based perception in autonomous driving under physical-world attacks,'' in \emph{2021 IEEE symposium on security and privacy (SP)}.\hskip 1em plus 0.5em minus 0.4em\relax IEEE, 2021, pp. 176--194.

\bibitem{yang2021robust}
K.~Yang, T.~Tsai, H.~Yu, M.~Panoff, T.-Y. Ho, and Y.~Jin, ``Robust roadside physical adversarial attack against deep learning in lidar perception modules,'' in \emph{Proceedings of the 2021 ACM Asia Conference on Computer and Communications Security}, 2021, pp. 349--362.

\bibitem{damodaran2023experimental}
D.~Damodaran, S.~Mozaffari, S.~Alirezaee, and M.~J. Ahamed, ``Experimental analysis of the behavior of mirror-like objects in lidar-based robot navigation,'' \emph{Applied Sciences}, vol.~13, no.~5, p. 2908, 2023.

\bibitem{vega2024slam2ref}
M.~A. Vega-Torres, A.~Braun, and A.~Borrmann, ``Slam2ref: advancing long-term mapping with 3d lidar and reference map integration for precise 6-dof trajectory estimation and map extension,'' \emph{Construction Robotics}, vol.~8, no.~2, p.~13, 2024.

\bibitem{koch2017detection}
R.~Koch, S.~May, and A.~N{\"u}chter, ``Detection and purging of specular reflective and transparent object influences in 3d range measurements,'' \emph{The International Archives of the Photogrammetry, Remote Sensing and Spatial Information Sciences}, vol.~42, pp. 377--384, 2017.

\bibitem{henley2023detection}
C.~Henley, S.~Somasundaram, J.~Hollmann, and R.~Raskar, ``Detection and mapping of specular surfaces using multibounce lidar returns,'' \emph{Optics Express}, vol.~31, no.~4, pp. 6370--6388, 2023.

\bibitem{kobayashi2024wip}
R.~Kobayashi, K.~Nomoto, Y.~Tanaka, G.~Tsuruoka, and T.~Mori, ``Wip: Shadow hack: Adversarial shadow attack against lidar object detection,'' in \emph{Proc. Symp. Veh. Security Privacy (VehicleSec)}, 2024, pp. 1--7.

\bibitem{kobayashiinvisible}
------, ``Invisible but detected: Physical adversarial shadow attack and defense on lidar object detection.''

\bibitem{zhu2024ae}
S.~Zhu, Y.~Zhao, K.~Chen, B.~Wang, H.~Ma, and C.~Wei, ``$\{$AE-Morpher$\}$: Improve physical robustness of adversarial objects against $\{$LiDAR-based$\}$ detectors via object reconstruction,'' in \emph{33rd USENIX Security Symposium (USENIX Security 24)}, 2024, pp. 7339--7356.

\bibitem{cao2019adversarial}
Y.~Cao, C.~Xiao, B.~Cyr, Y.~Zhou, W.~Park, S.~Rampazzi, Q.~A. Chen, K.~Fu, and Z.~M. Mao, ``Adversarial sensor attack on lidar-based perception in autonomous driving,'' in \emph{Proceedings of the 2019 ACM SIGSAC conference on computer and communications security}, 2019, pp. 2267--2281.

\bibitem{sun2020towards}
J.~Sun, Y.~Cao, Q.~A. Chen, and Z.~M. Mao, ``Towards robust $\{$LiDAR-based$\}$ perception in autonomous driving: General black-box adversarial sensor attack and countermeasures,'' in \emph{29th USENIX Security Symposium (USENIX Security 20)}, 2020, pp. 877--894.

\bibitem{zhu2021can}
Y.~Zhu, C.~Miao, T.~Zheng, F.~Hajiaghajani, L.~Su, and C.~Qiao, ``Can we use arbitrary objects to attack lidar perception in autonomous driving?'' in \emph{Proceedings of the 2021 ACM SIGSAC Conference on Computer and Communications Security}, 2021, pp. 1945--1960.

\bibitem{manivasagam2020lidarsim}
S.~Manivasagam, S.~Wang, K.~Wong, W.~Zeng, M.~Sazanovich, S.~Tan, B.~Yang, W.-C. Ma, and R.~Urtasun, ``Lidarsim: Realistic lidar simulation by leveraging the real world,'' in \emph{Proceedings of the IEEE/CVF Conference on Computer Vision and Pattern Recognition}, 2020, pp. 11\,167--11\,176.

\bibitem{manivasagam2023towards}
S.~Manivasagam, I.~A. B{\^a}rsan, J.~Wang, Z.~Yang, and R.~Urtasun, ``Towards zero domain gap: A comprehensive study of realistic lidar simulation for autonomy testing,'' in \emph{Proceedings of the IEEE/CVF International Conference on Computer Vision}, 2023, pp. 8272--8282.

\bibitem{sato2025realism}
T.~Sato, R.~Suzuki, Y.~Hayakawa, K.~Ikeda, O.~Sako, R.~Nagata, R.~Yoshida, Q.~A. Chen, and K.~Yoshioka, ``On the realism of lidar spoofing attacks against autonomous driving vehicle at high speed and long distance,'' in \emph{Proceedings of the Network and Distributed System Security Symposium (NDSS)}, 2025.

\bibitem{ouster2024os1}
\BIBentryALTinterwordspacing
Ouster, ``Os1 lidar sensor,'' 2024, accessed: 2024-07-09. [Online]. Available: \url{https://ouster.com/products/hardware/os1-lidar-sensor}
\BIBentrySTDinterwordspacing

\bibitem{dosovitskiy2017carla}
A.~Dosovitskiy, G.~Ros, F.~Codevilla, A.~Lopez, and V.~Koltun, ``{CARLA: An Open Urban Driving Simulator},'' in \emph{Proceedings of the 1st Annual Conference on Robot Learning}, 2017, pp. 1--16.

\bibitem{yin2021centerpoint}
T.~Yin, X.~Zhou, and P.~Krahenbuhl, ``Centerpoint: Center-based 3d object detection and tracking,'' in \emph{Proceedings of the IEEE/CVF Conference on Computer Vision and Pattern Recognition (CVPR)}, 2021, pp. 11\,784--11\,793.

\bibitem{fukunagarandom}
M.~Fukunaga and T.~Sugawara, ``Random spoofing attack against {LiDAR}-based scan matching {SLAM}.''

\bibitem{hau2021object}
Z.~Hau, K.~T. Co, S.~Demetriou, and E.~C. Lupu, ``Object removal attacks on lidar-based {3d} object detectors,'' \emph{arXiv preprint arXiv:2102.03722}, 2021.

\bibitem{tibebu2021lidar}
H.~Tibebu, J.~Roche, V.~De~Silva, and A.~Kondoz, ``Lidar-based glass detection for improved occupancy grid mapping,'' \emph{Sensors}, vol.~21, no.~7, p. 2263, 2021.

\bibitem{yang2024evaluating}
B.~Yang, T.~M. Triet~Pham, and J.~Yang, ``Evaluating and improving the robustness of lidar-based localization and mapping,'' \emph{arXiv e-prints}, pp. arXiv--2409, 2024.

\bibitem{piroli2023towards}
A.~Piroli, V.~Dallabetta, J.~Kopp, M.~Walessa, D.~Meissner, and K.~Dietmayer, ``Towards robust 3d object detection in rainy conditions,'' in \emph{2023 IEEE 26th International Conference on Intelligent Transportation Systems (ITSC)}.\hskip 1em plus 0.5em minus 0.4em\relax IEEE, 2023, pp. 3471--3477.

\bibitem{rivero2021effect}
J.~R.~V. Rivero, T.~Gerbich, B.~Buschardt, and J.~Chen, ``The effect of spray water on an automotive lidar sensor: A real-time simulation study,'' \emph{IEEE Transactions on Intelligent Vehicles}, vol.~7, no.~1, pp. 57--72, 2021.

\bibitem{dong2023benchmarking}
Y.~Dong, C.~Kang, J.~Zhang, Z.~Zhu, Y.~Wang, X.~Yang, H.~Su, X.~Wei, and J.~Zhu, ``Benchmarking robustness of 3d object detection to common corruptions,'' in \emph{Proceedings of the IEEE/CVF Conference on Computer Vision and Pattern Recognition}, 2023, pp. 1022--1032.

\bibitem{cao2023you}
Y.~Cao, S.~H. Bhupathiraju, P.~Naghavi, T.~Sugawara, Z.~M. Mao, and S.~Rampazzi, ``You can't see me: Physical removal attacks on $\{$lidar-based$\}$ autonomous vehicles driving frameworks,'' in \emph{32nd USENIX Security Symposium (USENIX Security 23)}, 2023, pp. 2993--3010.

\bibitem{hau2021shadow}
Z.~Hau, S.~Demetriou, L.~Mu{\~n}oz-Gonz{\'a}lez, and E.~C. Lupu, ``Shadow-catcher: Looking into shadows to detect ghost objects in autonomous vehicle 3d sensing,'' in \emph{Computer Security--ESORICS 2021: 26th European Symposium on Research in Computer Security, Darmstadt, Germany, October 4--8, 2021, Proceedings, Part I 26}.\hskip 1em plus 0.5em minus 0.4em\relax Springer, 2021, pp. 691--711.

\bibitem{hallyburton2022security}
R.~S. Hallyburton, Y.~Liu, Y.~Cao, Z.~M. Mao, and M.~Pajic, ``Security analysis of $\{$Camera-LiDAR$\}$ fusion against $\{$Black-Box$\}$ attacks on autonomous vehicles,'' in \emph{31st USENIX Security Symposium (USENIX Security 22)}, 2022, pp. 1903--1920.

\end{thebibliography}

\begin{appendices}
\section{Empirical Model Validation and Parameter Analysis}
\subsection{Fitted Model Parameters and Physical Interpretation}
\label{app:model_params}

This appendix provides the specific parameter values for the empirical models developed in this study, as determined by fitting the model equations to our comprehensive experimental dataset. A subsequent glossary (Table~\ref{tab:parameter_glossary}) offers a detailed physical interpretation of each parameter.

\noindent \tikz[baseline=(char.base)]\node[shape=circle,draw,inner sep=0.8pt] (char) {1}; \textbf{Fitted Parameter Values.}~The final coefficients for each of the four empirical models are presented in Table~\ref{tab:fitted_values}. These values were obtained through a global curve-fitting process across all relevant experimental configurations and represent the parameters used to generate the ``Fitted Model" lines in our results.
\begin{table}[ht]\small
\centering
\caption{Fitted Values for Empirical Model Parameters}
\label{tab:fitted_values}
\renewcommand{\arraystretch}{1.1}
\resizebox{\linewidth}{!}{%
\begin{tabular}{|l|c|r|}
\hline
\textbf{Model} & \textbf{Parameter} & \textbf{Fitted Value} \\
\hline
\hline
\textbf{Artifact Point Count} ($N_{artifact}$) & $c_0$ & 2500.0 \\
($R^2 \approx 0.85$ global) & $\beta$ & 1.25 \\
& $\gamma$ & 3.5 \\
& $\mu$ & 3.0 m \\
& $\sigma$ & 1.5 m \\
\hline
\textbf{Appearance Probability} ($P_{appearance}$) & $k$ & 15.0 (fixed) \\
(Predictive Model) & $b_0^{\min}$ & -2.858 \\
& $b_1, c_1$ & 0.1389 \\
& $b_2$ & -0.3003 \\
& $b_0^{\max}$ & -0.946 \\
& $c_2$ & 1.5390 \\
\hline
\textbf{Lateral Offset} ($X_{artifact}$) & $c_X$ & 0.98 \\
($R^2 \approx 0.98$ global) & $\delta_X$ & -0.05 m \\
\hline
\textbf{Radial Distance} ($R_{artifact}$) & $c_R$ & 1.02 \\
($R^2 \approx 0.92$ global) & $n_d$ & 0.88 \\
& $a_1$ & 0.15 \\
& $a_2$ & 0.08 \\
\hline
\end{tabular}
}
\end{table}


\noindent \tikz[baseline=(char.base)]\node[shape=circle,draw,inner sep=0.8pt] (char) {2}; \textbf{Glossary of Model Parameters.}~Table~\ref{tab:parameter_glossary} provides a detailed physical interpretation for each parameter listed above, connecting its mathematical role to the underlying geometric and physical phenomena observed during the LiDAR-mirror interactions.

\begin{table*}[h!]\small
\centering
\caption{Physical Interpretation of Model Parameters}
\label{tab:parameter_glossary}
\renewcommand{\arraystretch}{1.1} 
\begin{tabular}{|c|p{0.75\textwidth}|}
\hline
\textbf{Parameter} & \textbf{Description and Physical Interpretation} \\
\hline
\hline
$c_0$ & \textbf{Amplitude Scaling Factor}: Represents the baseline maximum number of points generated under ideal conditions (e.g., perfect alignment, unit area, and zero angle). \\
\hline
$\beta$ & \textbf{Area Scaling Exponent}: Controls how sensitively the point count scales with the mirror's surface area ($A$). A value of $\beta > 1$ suggests a super-linear relationship, possibly due to beam uniformity effects. \\
\hline
$\gamma$ & \textbf{Angle Foreshortening Exponent}: Models the reduction in effective area due to the mirror's tilt angle, related to the $\cos(\theta)$ term. \\
\hline
$\mu, \sigma$ & \textbf{Effective Distance Window}: The mean ($\mu$) and standard deviation ($\sigma$) of the Gaussian term, defining the optimal distance for point generation and the range over which it occurs. \\
\hline
\hline
$k$ & \textbf{Transition Steepness}: A fixed constant controlling the sharpness of the probability window's edges. A high value creates a near-instantaneous transition from $P=0$ to $P=1$. \\
\hline
$d_{\min}, d_{\max}$ & \textbf{Appearance Window Thresholds}: The calculated minimum and maximum distances for a valid reflection path, determined by the linear expressions involving coefficients $b_0^{\min}, b_1, b_2$ and $b_0^{\max}, c_1, c_2$ respectively.. These define the ``window of opportunity" for the attack. \\
\hline
\hline
$c_X$ & \textbf{Geometric Scaling Factor}: An empirical coefficient scaling the trigonometric model ($d \tan(2\theta)$). A value close to 1.0, like our fitted 0.98, confirms the high accuracy of the underlying physics model. \\
\hline
$\delta_X$ & \textbf{System Bias}: A constant offset representing any minor, systematic physical misalignment in the experimental setup. \\
\hline
\hline
$c_R$ & \textbf{Base Distance Factor}: A primary scaling coefficient for the perceived radial distance. A value near 1.0 indicates the model is well-calibrated. \\
\hline
$n_d$ & \textbf{Distance Power-Law Exponent}: Captures the non-linear growth of perceived distance. Our fitted value of $0.88~(< 1)$ confirms a sub-linear relationship, likely due to second-order effects like beam spillover. \\
\hline
$a_1, a_2$ & \textbf{Angular Offset Coefficients}: Parameters of the polynomial that models the additive offset to the radial distance caused by the mirror's tilt angle. \\
\hline
\end{tabular}
\end{table*}

\subsection{Validation of Models Across Experimental Configurations}

\begin{figure}[t]
\centering
    \includegraphics[width=\linewidth]{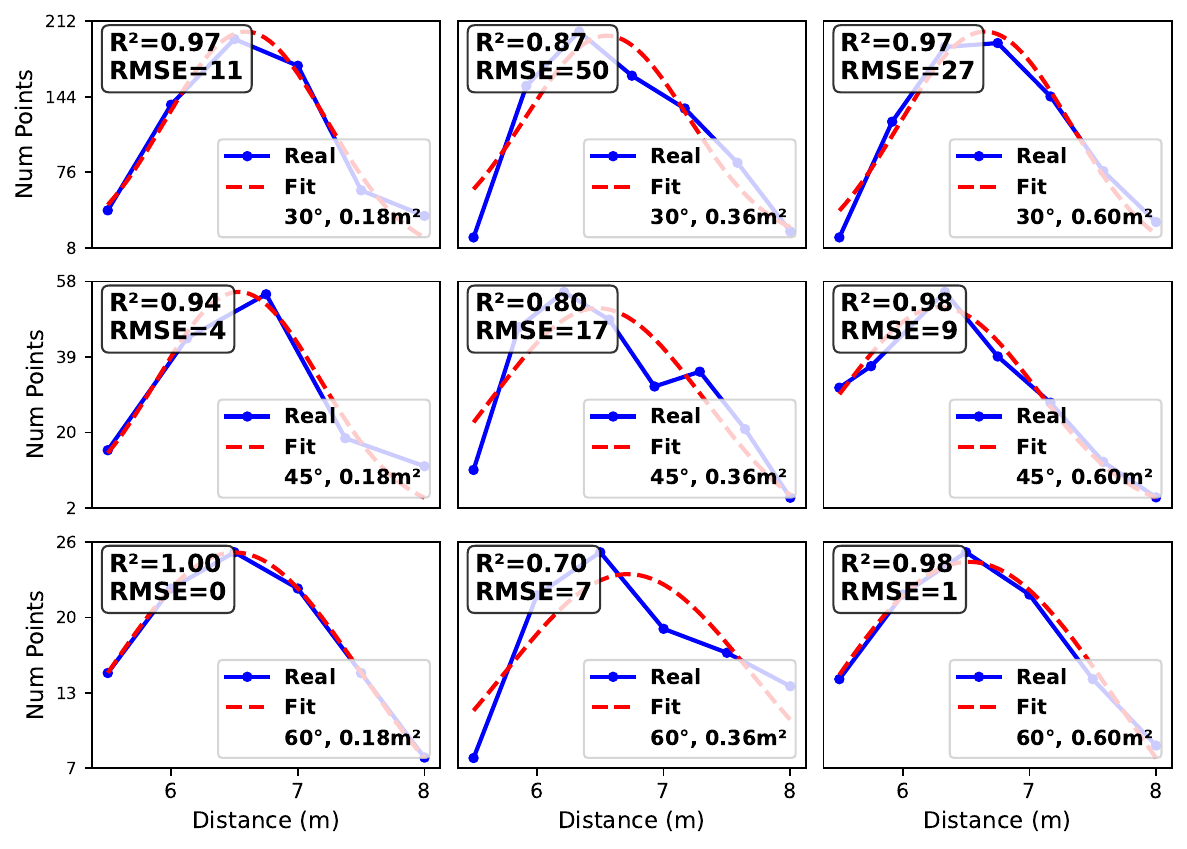}
    \captionsetup{font=small}
\caption{Validation of the analytical model for predicting phantom point cloud size. Each subplot compares the experimentally measured number of artifact points ('Real', solid blue line) with the predictions from our analytical model ('Fit', dashed red line) as a function of the vehicle's distance from the mirror for different $A$ and $\theta$. The high R$^2$ values and low RMSE demonstrate that the model accurately captures the trend of point cloud intensity across a wide range of configurations.}
 \label{figure: fit_Number_point}
\end{figure}

\noindent \tikz[baseline=(char.base)]\node[shape=circle, fill=black, inner sep=0.8pt, text=white] (char) {1}; \textbf{Model (a): Artifact Point Count ($N_{artifact}$).}~The Gaussian-like model for the number of artifact points demonstrates a strong predictive capability across all tested configurations, with R² values ranging from 0.70 to a near-perfect 1.00, as shown in Figure~\ref{figure: fit_Number_point}. This high correlation indicates that the model successfully captures the fundamental trend of how point cloud density evolves as the vehicle approaches the mirror. The model accurately represents the characteristic rise-and-fall pattern of point counts, which is governed by the geometric interplay between the LiDAR's conical beam spread and the mirror's fixed surface area. When the vehicle is far away, the beam's cross-section is large, but only a small portion intersects the mirror, yielding few points. As the vehicle approaches, the solid angle of the beam subtended by the mirror increases, causing the point count to rise sharply to a peak at an optimal alignment distance. Beyond this peak, the vehicle begins to move past the mirror's field of view, causing the alignment to degrade and the point count to drop.

\begin{figure}[ht]
\centering
    \includegraphics[width=\linewidth]{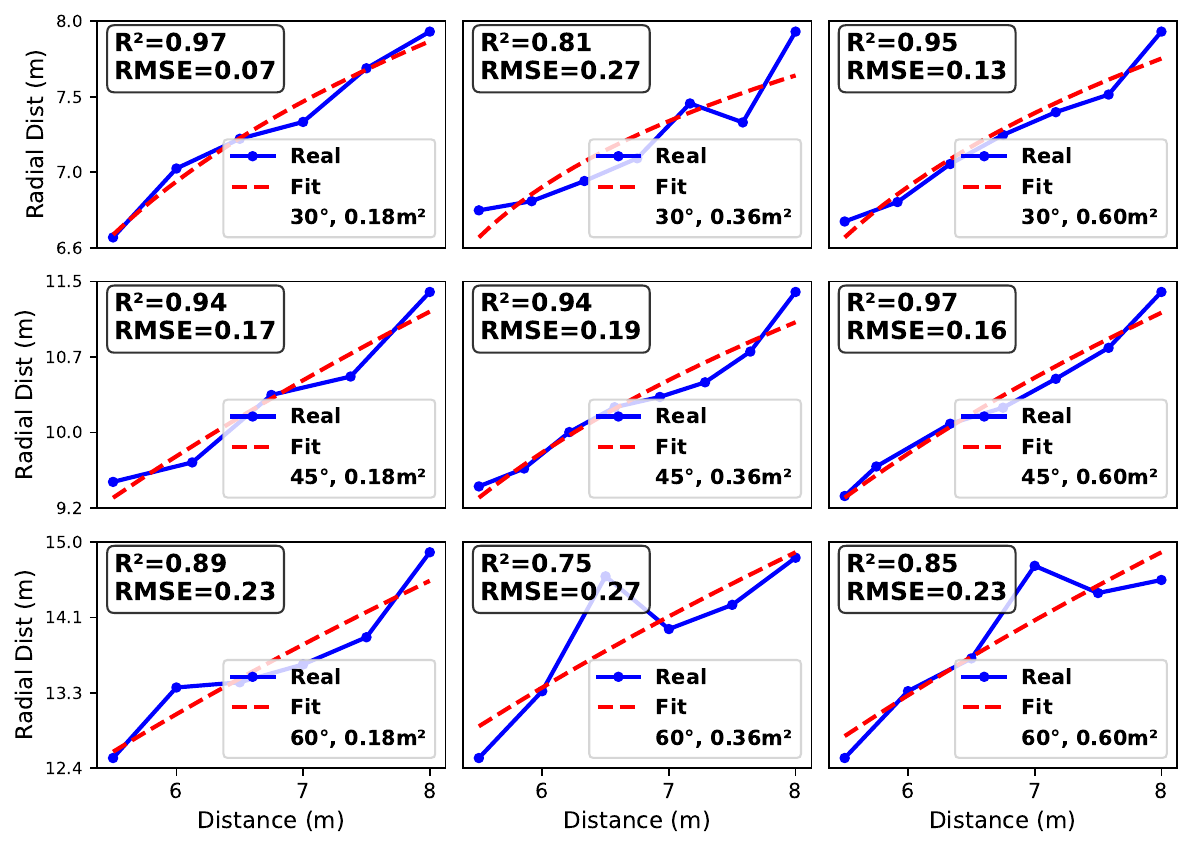}
    \captionsetup{font=small}
\caption{Model validation for the predicted radial distance of the phantom artifact. The plots compare measured data ('Real') against model predictions ('Fit') across three tilt angles (rows) and three surface areas (columns). The model accurately tracks the near-linear relationship between the vehicle's real distance and the phantom's perceived radial distance, with the high R$^2$ values confirming its predictive accuracy for the attack's ranging geometry.}
 \label{figure: fit_radial_distance}
\end{figure}

The mirror's surface area ($A$) emerges as the most dominant factor in determining the \textit{magnitude} of the virtual object. As shown in Figure~\ref{figure: fit_radial_distance}, increasing the area from 0.18\,m$^2$ to 0.60\,m$^2$ causes a dramatic increase in the peak number of points (e.g., from a maximum of 25 to over 400 at a 30\textdegree{} angle). This is intuitively explained by the mirror acting as a larger ``net," intercepting a significantly greater fraction of the LiDAR's emitted pulses and thus generating a denser, more cohesive virtual object. Conversely, as the tilt angle ($\theta$) increases from 30\textdegree{} to 60\textdegree{}, the peak number of artifact points noticeably decreases. This is attributable to two geometric effects: first, the foreshortening of the mirror, which reduces its effective cross-sectional area as perceived by the LiDAR beam (proportional to $A \cos\theta$), and second, a potentially less efficient reflection path from the secondary object at steeper angles.

\begin{figure}[ht]
\centering
    \includegraphics[width=\linewidth]{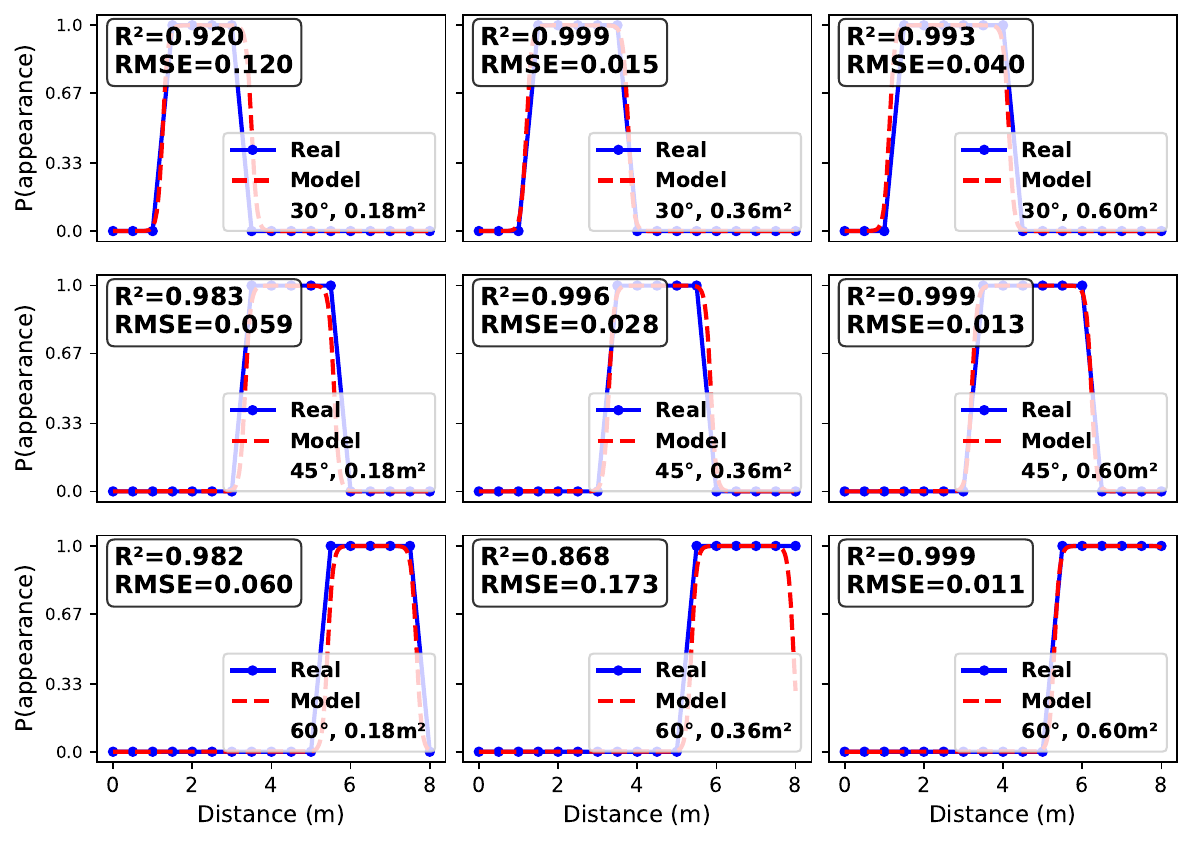}
    \captionsetup{font=small}
\caption{Validation of the analytical model for predicting the probability of phantom point cloud appearance as a function of distance. Each subplot compares the experimentally measured probability ('Real') with the model's prediction ('Model'). The extremely high R$^2$ values and low RMSE across all nine configurations demonstrate the model's high accuracy in predicting the precise distance window where the phantom becomes visible.}
 \label{figure: fit_probability}
\end{figure}

\noindent \tikz[baseline=(char.base)]\node[shape=circle, fill=black, inner sep=0.8pt, text=white] (char) {2}; \textbf{Model (b): Appearance Probability ($P_{app}$).}~The double-sigmoid model provides an exceptionally accurate fit for the appearance probability, with R² values consistently above 0.87 and often approaching 1.00 (see Figure~\ref{figure: fit_probability}). The sharp, window-like transitions in the experimental data are captured almost perfectly by the model, validating its underlying formulation as a product of two logistic functions. This model confirms that an artifact's appearance is not continuous but exists within a finite ``window of opportunity" defined by a minimum and maximum distance ($d_{\min}$, $d_{\max}$). Outside this window, a valid geometric reflection path between the LiDAR, mirror, and secondary object does not exist, and the probability is effectively zero.

The analysis reveals that a larger mirror area ($A$) directly translates to a \textit{wider} probability window. This is because a bigger mirror offers greater geometric tolerance, allowing a valid reflection path to be established earlier and maintained for a longer duration as the vehicle passes. From an attacker's perspective, this means a larger mirror makes the attack effective over a greater range of distances. The tilt angle ($\theta$), in contrast, primarily influences the \textit{position} of this window along the distance axis. By changing the overall geometry of the reflection, different angles cause the $d_{\min}$ and $d_{\max}$ thresholds to shift, highlighting a key parameter an adversary can tune to control precisely when and where an artifact becomes visible.

\begin{figure}[t]
\centering
    \includegraphics[width=\linewidth]{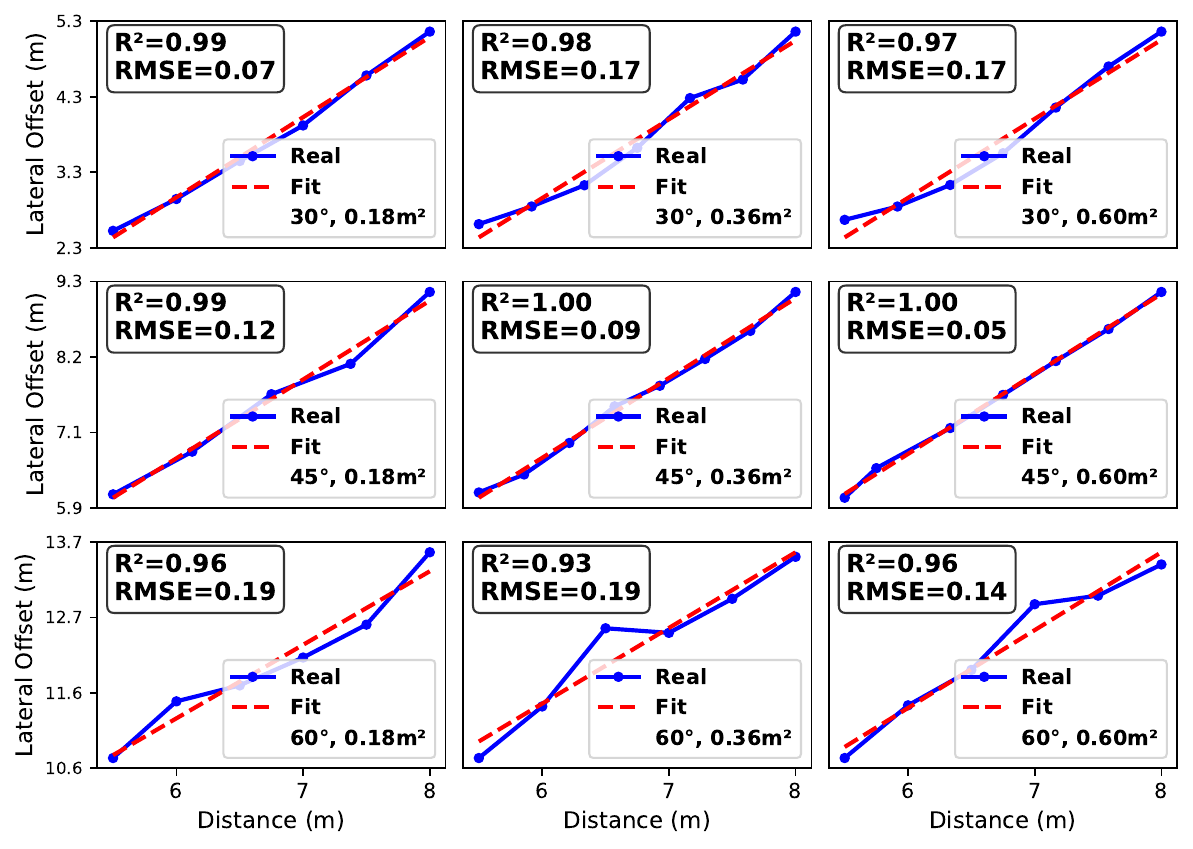}
    \captionsetup{font=small}
\caption{Model validation for the predicted lateral position of the phantom artifact. The plots compare measured data ('Real') against model predictions ('Fit') across three tilt angles (rows) and three surface areas (columns). The model accurately tracks the near-linear relationship between vehicle distance and the phantom's lateral offset, with high R$^2$ values confirming its predictive accuracy for the attack's spatial geometry.}
 \label{figure: fit_Lateral_offset}
\end{figure}

\noindent \tikz[baseline=(char.base)]\node[shape=circle, fill=black, inner sep=0.8pt, text=white] (char) {3}; \textbf{Model (c): Lateral Offset ($X_{artifact}$).}~The linear model for the lateral offset of the virtual artifact demonstrates a near-perfect fit, with R² values consistently at 0.93 or higher and often exceeding 0.99 (see Figure~\ref{figure: fit_Lateral_offset}). This exceptionally strong correlation confirms that the model accurately reflects the fundamental trigonometry of specular reflection. The results show that the lateral offset of the virtual point's centroid, $X_{artifact}$, increases linearly with the LiDAR-to-mirror distance, $d$. This is a direct consequence of geometric scaling: the LiDAR, mirror, and virtual point form a triangle that scales proportionally as the distance $d$ increases.

The mirror's tilt angle ($\theta$) is the most influential factor on the \textit{magnitude} of the lateral offset, governing the slope of the linear relationship. As the angle increases from 30\textdegree{} to 60\textdegree{}, the y-axis scales show a dramatic increase in the offset, a behavior precisely predicted by the $\tan(2\theta)$ term in the underlying physical model. This demonstrates that an attacker can use a steeper mirror angle to project a virtual object much further to the side, even from a close distance. Critically, the mirror's surface area ($A$) has no discernible effect on the artifact's location. The position of a virtual point is dictated purely by the angles of reflection, not by the size of the surface that causes it. This leads to a crucial insight: an attacker uses mirror \textbf{angle} to control the \textbf{placement} of an artifact, while using mirror \textbf{area} to control its \textbf{believability}.

\noindent \tikz[baseline=(char.base)]\node[shape=circle, fill=black, inner sep=0.8pt, text=white] (char) {4}; \textbf{Model (d): Radial Distance ($R_{artifact}$).}~The power-law model for the perceived radial distance of the artifact also provides a very strong fit, with R² values consistently high (mostly above 0.85), as shown in Figure~\ref{figure: fit_radial_distance}. This confirms that the model accurately captures the relationship between the vehicle's actual position and the perceived distance of the virtual artifact. As the actual LiDAR-to-mirror distance ($d$) increases, the perceived radial distance ($R_{artifact}$) of the virtual point increases in a nearly linear fashion. The power-law formulation ($d^{n_d}$) is slightly more accurate than a pure linear model because the total folded path length ($d_{LM} + d_{MS}$) is a complex function of $d$.

The tilt angle ($\theta$) significantly affects the overall magnitude of the perceived distance. As the angle increases from 30\textdegree{} to 60\textdegree{}, the entire set of radial distance measurements shifts upwards to higher values. This is because a steeper angle creates a longer reflection path from the mirror to the secondary object ($d_{MS}$), which adds a substantial offset to the total path length measured by the LiDAR. Similar to the lateral offset, the mirror's surface area ($A$) has no significant impact on the perceived radial distance. This reinforces the conclusion that the artifact's location in 3D space is a function of geometry (angle) and path length (distance), not the size of the mirror. An attacker's choice of mirror size controls the artifact's density, not its perceived position.

\section{Supplementary Point Cloud Visualizations}
\label{app:point_clouds}
This appendix presents supplementary visualizations of the raw LiDAR point clouds generated under different OAA and ORA attack These figures provide direct visual evidence of how key physical parameters influence the raw sensor data, leading to the perception failures discussed in the main paper.
\subsection{OAA: Phantom Point Cloud Characteristics}
\label{app:point_clouds_OAA}
The following analysis details how the physical properties of the mirror setup in an OAA scenario control the characteristics of the resulting phantom point cloud.

\begin{figure}[!ht]
    \centering
    \includegraphics[width=\linewidth]{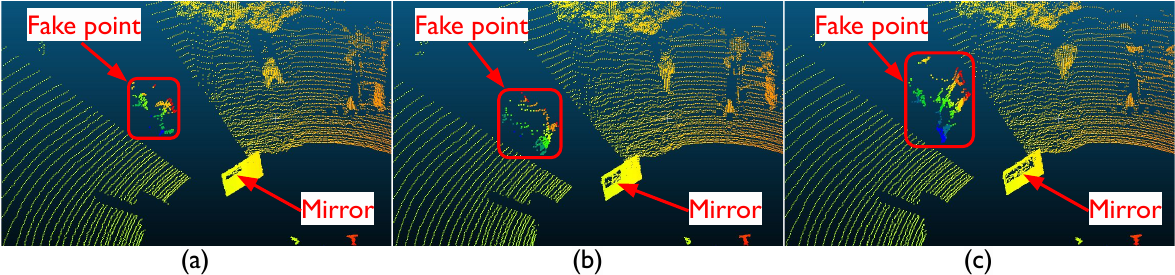}
    \captionsetup{font=small}
    \caption{Impact of mirror surface area on phantom point cloud characteristics: \textbf{(a)} Two tiles (0.18\,m$^2$); \textbf{(b)} Four tiles (0.36\,m$^2$); \textbf{(c)} Six tiles (0.60\,m$^2$).}
    \label{fig:OAAappendix_size}
\end{figure}

\noindent \tikz[baseline=(char.base)]\node[shape=circle,draw,inner sep=0.8pt] (char) {1}; \textbf{Impact of Mirror Surface Area.}~Figure~\ref{fig:OAAappendix_size} illustrates the critical role of mirror surface area in determining the quality and believability of the phantom point cloud. The vehicle and mirror positions were held constant while only the number of mirror tiles was varied.

\noindent \textbf{Discussion.}~The surface area of the mirror directly dictates the severity and confidence of the perceptual deception.
Even with 2 tiles, the attack demonstrates initial success. While the resulting point cloud is sparse and fragmented, it forms a sufficiently dense local cluster to surpass basic noise filters, leading Autoware to classify it as an 'UNKNOWN' object. This finding is critical, as it proves that a tangible perception-level impact can be achieved with a minimal adversarial setup.
Increasing the area to four tiles marks a significant escalation of the attack. The point cloud becomes substantially denser, acquiring a volumetric presence that begins to mimic a real-world object. This added structure is sufficient to deceive the detector's learned features, causing a semantic misclassification where the phantom is now identified as a 'CAR', with 65\% confidence.
The six-tile configuration produces a high-fidelity deception. It generates a dense, spatially cohesive point cloud that forms a well-defined planar structure. This provides a convincing match to the detector's model priors, solidifying the misclassification as a 'CAR' with high confidence (74\%). This progression clearly demonstrates that an attacker can tune the severity of the deception: from triggering a low-level anomaly detection ('UNKNOWN') to inducing a confident, semantic misclassification ('CAR') simply by increasing the reflective surface area.

\begin{figure}[!ht]
    \centering
    \includegraphics[width=\linewidth]{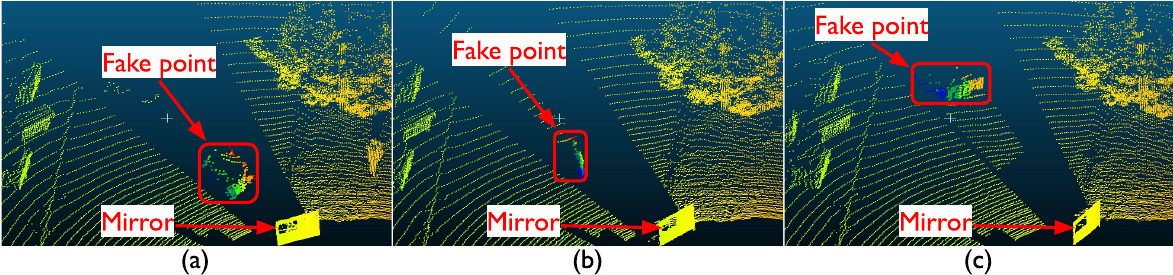}
    \captionsetup{font=small}
    \caption{Impact of mirror tilt angle on the perceived location of the OAA phantom: \textbf{(a)} 30$^\circ$ angle; \textbf{(b)} 45$^\circ$ angle; \textbf{(c)} 60$^\circ$ angle.}
    \label{fig:OAAappendix_angle}
\end{figure}

\noindent \tikz[baseline=(char.base)]\node[shape=circle,draw,inner sep=0.8pt] (char) {2}; \textbf{Impact of Mirror Tilt Angle.}~Figure~\ref{fig:OAAappendix_angle} demonstrates how the mirror's tilt angle dictates the spatio-temporal dynamics of the phantom—its perceived distance, spatial position, and timing.

\noindent \textbf{Discussion.}~The mirror's tilt angle directly governs the phantom's spatio-temporal dynamics, controlling its perceived location, distance, and the timing of its appearance.
At a shallow 30$^\circ$ angle, the stringent geometric requirements for reflection result in a ``late-onset" phantom. It appears abruptly only when the vehicle is at close physical range. This geometry induces a significant lateral shift, positioning the phantom at a short perceived distance and deep inside the arc of the turn. The narrow alignment window makes its appearance brief and transient.

Conversely, a steep 60$^\circ$ angle creates a ``persistent barrier" phantom. The wide cone of effective reflection allows for early detection in the vehicle's approach. It materializes at a greater perceived distance and, due to minimal lateral shift, is projected far down the vehicle's forward path. This configuration provides a long-lived, stable reflection that remains in the sensor's field of view for a significant portion of the trajectory. The intermediate 45$^\circ$ angle exhibits transitional properties but is notable for producing significant ``flicker" in the raw data—a frame-to-frame instability in both the point cloud's density and its precise perceived location. This highlights how the attack geometry can be tuned not just for position, but also for consistency.

\subsection{ORA: Robustness of Obstacle Removal Across Angles}
\label{app:point_clouds_ORA}
The following visualizations for the ORA scenario demonstrate the attack's effectiveness and robustness. The primary goal is to show that even at various, non-ideal mirror angles, the attack successfully removes the real obstacle from the LiDAR's field of view, creating the dangerous illusion of a clear path.
\begin{figure}[ht]
\centering
    \includegraphics[width=\linewidth]{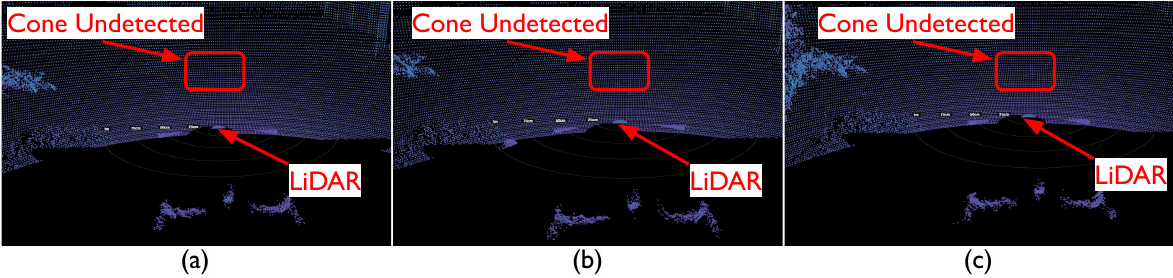}
    \captionsetup{font=small}
    \caption{LiDAR point clouds under ORA (object removal) at various mirror angles. \textbf{(a)} 15\textdegree{} mirror tilt – the traffic cone in front of the car yields no LiDAR returns; the cone is effectively invisible (no points in its position). \textbf{(b)} 30\textdegree{} – again, the cone is fully undetected. \textbf{(c)} 45\textdegree{} – even at this steeper angle, no returns from the cone are seen.}
    \label{PC:OORA_angles-appendix}
\end{figure}

\noindent \textbf{Discussion.}~The visualizations in Figure~\ref{PC:OORA_angles-appendix} demonstrate the insidious effectiveness and robustness of the ORA attack. The baseline scan (top-left) provides the ground truth, clearly showing the point cloud of the real traffic cone that the perception system would normally identify as an obstacle.
The subsequent images show the result of introducing the mirror at three different tilt angles: 15$^\circ$, 30$^\circ$, and 45$^\circ$. In all three attack configurations, the outcome is absolute and identical: the points corresponding to the traffic cone are completely absent from the LiDAR scan. Critically, this is not a simple data void or sensor dropout. The LiDAR beams that would have struck the cone are instead deflected by the mirror onto the ground plane. The sensor receives these valid-looking returns from the road surface and, as a result, perceives a continuous, navigable path that extends directly through the physical space occupied by the real obstacle.
The consistency of this effect across the tested angles of 15$^\circ$, 30$^\circ$, and 45$^\circ$ demonstrates the attack's high degree of robustness. It does not require fragile, precise alignment to succeed, giving it a wide operational envelope and making it a practical and realistic threat. This active replacement of true object data with plausible, but false, ground data makes the ORA exceptionally dangerous, as it is designed to silently deceive the perception system into a state of high confidence about a clear path that is, in reality, obstructed.
\end{appendices}

\end{document}